%% file: Main.tex
\documentclass[aps,prl,reprint]{revtex4-2}
\usepackage{xcolor}
\usepackage{graphicx}
\usepackage{float}
\usepackage{dcolumn}
\usepackage{bm}
\usepackage[english]{babel}

\begin{document}

\preprint{APS/123-QED}

\title{Quadratic Supercontinuum Generation from UV to Mid-IR in Lithium Niobate Nanophotonics}

\author{Selina Zhou$^{*}$}

\author{Maximilian Shen$^{*}$}%

\author{Ryoto Sekine$^{*}$}

\author{Nicolas Englebert$^{*}$}

\author{Thomas Zacharias}

\author{Benjamin Gutierrez}

\author{Robert M. Gray}

\author{Justin Widjaja}

\author{Alireza Marandi$^{\dagger}$}

\affiliation{%
 Department of Electrical Engineering, California Institute of Technology\\
 $^{*}$These authors contributed equally to this work \\
 $^{\dagger}$marandi@caltech.edu
}%


\date{\today}

\begin{abstract}

Supercontinuum light sources are widely used for applications ranging from imaging to sensing and frequency comb stabilization. The most common mechanisms for their generation rely on cubic nonlinearities, for instance in crystals, optical fibers, and integrated photonics. However, quadratic supercontinuum generation (QSCG) offers potential for enhanced energy efficiency and broader spectral coverage because of the typically much stronger nonlinearity and ability to achieve both coherent up- and down-conversion via three-wave mixing processes. Despite such potentials, demonstrations of quadratic supercontinuum generation in integrated photonic waveguides have been sparse and have barely surpassed their cubic counterparts in terms of spectral coverage and energy-efficiency. Here, we introduce a new dispersion engineering principle and experimentally demonstrate purely quadratic supercontinuum generation in lithium niobate nano-waveguides substantially outperforming previous demonstrations in integrated photonics. In one device, by engineering a near-zero dispersion profile and using a single poling period for quasi-phase matched saturated second-harmonic generation, we achieve robust and energy efficient multi-octave QSCG in lithium niobate nano-waveguide with only femtojoules of pump pulse energy, which is a few orders of magnitude less energy compared to other comparable spectra achieved in single-pass waveguides. In another device, we use a flat dispersion profile with two distant zero crossings of group velocity dispersion (GVD) to achieve broadband difference-frequency generation (DFG) for extending the spectral coverage further into the mid-IR and cover the entire transparency window of lithium niobate from 350 nm to 5 µm, which corresponds to 3.8 octaves, surpassing previous SCG demonstrations in waveguides. Our results showcase how DFG-assisted QSCG can access hard-to-access spectral regions, especially the UV, visible, and mid-infrared, in an energy-efficient fashion by properly utilizing dispersion engineering and quasi-phase matching.

\end{abstract}

\maketitle

In the past decade, the development of broadband on-chip frequency comb sources has attracted a large amount of attention\,\cite{kippenberg_dissipative_2018,pasquazi_micro-combs_2018,diddams_optical_2020}. Various integrated platforms have emerged over the years, demonstrating distinct behaviors in terms of spectral coverage, efficiency, and integration capabilities\,\cite{davenport_integrated_2018,gaeta_photonic-chip-based_2019,van_gasse_recent_2019,chang_integrated_2022,meng_dissipative_2022,kazakov_driven_2025}. For instance, silicon nitride (SiN) platforms have achieved notable success due to their low loss and strong third-order ($\chi^{(3)}$, Kerr) nonlinearity, demonstrating octave-spanning frequency combs \cite{Pfeiffer:17,Brasch2017}. More recently, thin-film lithium niobate (TFLN) has emerged as a promising platform for the realization of broadband and compact frequency comb sources, relying on strong second-order ($\chi^{(2)}$) nonlinear effects, a broad transparency window, and versatile electro-optic (EO) modulation capabilities\,\cite{Gaeta19_Review,Bres23_Review}.
These demonstrations have led to revolutions in numerous fields of applied science such as precision spectroscopy\,\cite{MGSuh16, Stern:2020, Bao2021}, optical frequency synthesis\,\cite{Singh20,Spencer2018}, optical links \cite{Marin-Palomo2017,Rizzo2023} and clocks\,\cite{newman_architecture_2019}, metrology \cite{Pupeza2021}, and biomedical sensing \cite{Yu:19, Marchand2021}.

One of the frequency comb sources of peculiar interest is the one based on supercontinuum generation (SCG), which can convert an ultrashort pulse into a phase-coherent and multi-octave spectrum via nonlinear interactions and dispersion engineering\,\cite{dudley_supercontinuum_2006,Dudley_2010}. Among others, they enable broadband spectroscopy, imaging, sensing, and self-referencing\,\cite{Yue2024IntegratedSCG}. So far, integrated SCG has mainly been achieved through third-order nonlinear processes\,\cite{Cao2022,Lu:19_chi3,Yu:19_chi3,gao2025_chi3}. These $\chi^{(3)}$-based SCG typically rely on self-phase modulation (SPM), cross-phase modulation (XPM), four-wave mixing (FWM) processes, and/or soliton fission\,\cite{Dudley_2010}, but also on dispersive waves emission to further enhance the spectral coverage\,\cite{Guo2018}. Yet, since the Kerr effect is inherently weak, these processes usually require high input pulse peak power, which can prohibit their use. 

\textit{Quadratic} SCGs (QSCG) aim to address this limitation as, through quasi-phase matching (QPM) techniques like periodic poling\,\cite{yamada_1993_qpm}, the magnitude of quadratic nonlinear effects can significantly exceed that of third-order nonlinear effects. The spectral broadening in QSCG mainly stems from second harmonic generation (SHG) and its saturation behavior\,\cite{Marc:23}, phase-mismatched SHG\,\cite{Phillips2011_chi2}, and/or other three-wave mixing processes such as difference-frequency generation and sum-frequency generation (SFG)\,\cite{Jankowski:22:quasi,Kowligy:18}. Because of this multitude of $\chi^{(2)}$ processes, dispersion engineering and QPM in quadratic nano-waveguides play key roles, \cite{Marc:23,Wu2024}, acting as extra degrees of freedom to precisely tailor the different nonlinear processes over a broad spectral range. For instance, careful control of group velocity dispersion (GVD) and group velocity mismatch (GVM) can further enhance spectral broadening efficiency and give access to the mid-IR spectral region\cite{Marc:23}. 

Despite these strengths and richness, most of the integrated demonstrations of purely quadratic SCG have so far been limited to saturated SHG\,\cite{Marc:23,Hamrouni:24}, enabled by dispersion engineering\,\cite{Jankowski2021,Ledezma:22} and barely surpassing the efficiency and bandwidth of their cubic counterparts. While some designs combining $\chi^{(2)}$ and $\chi^{(3)}$ nonlinearities have demonstrated potential for highly broadband comb generation in TFLN\,\cite{Wu2024, fan2025_chirpchi2}, they still require up to tens or hundreds of picojoules of pump pulse energy and show limited spectral content in the mid-IR, primarily due to the silica substrate absorption. This has motivated the use of alternative substrates such as sapphire to extend the TFLN transparency window into the mid-IR, enabling the formation of spectral content up to 4 \textmu m\,\cite{Mishra:21,Mishra:22}. Although TFLN-on-sapphire QSCGs relying on saturated SHG have been successfully demonstrated, their frequency content has so far been limited in the near-IR\,\cite{Hamrouni:24}. Yet, for many applications, such as sensing\,\cite{Yu:19, Marchand2021} and spectroscopy\,\cite{MGSuh16, Stern:2020, Bao2021}, access to mid-IR wavelengths is crucial. Interestingly, some theoretical works show that DFG can be advantageously combined with saturated SHG QSCG to access these elusive spectral regions\,\cite{Xiong24_cascadePPLN,Shen:25}. Yet, to the best of our knowledge, an experimental demonstration of such a DFG-assisted QCSG is still missing.

In this work, we experimentally demonstrate QSCG in three different dispersion-engineered lithium niobate nano-waveguides, substantially outperforming previous demonstrations in integrated photonics. 
In the first device, we engineer a near-zero dispersion profile and use a single poling period for quasi-phase-matched saturated SHG. We achieve robust and energy-efficient multi-octave QSCG in TFLN waveguides with only femtojoules of pump pulse energy, which is a few orders of magnitude less energy compared to other comparable spectra achieved in single-pass waveguides (see Supplementary Table S1). Using the second device, we show that while targeting low GVD and GVM can lead to efficient QSCG, as demonstrated with the first device, proper phase- and group velocity-matching of DFG processes offers a path to further extend the long-wavelength side of the generated spectrum towards the mid-IR. 
Specifically, we experimentally demonstrate such a DFG-assisted QCSG, and measure spectral content beyond 4 \textmu m. 
Finally, using our third device with a sapphire substrate, we demonstrate 3.8 octaves DFG-assisted QCSG, filling the entire TFLN transparency window and surpassing previous QSCG demonstrations in waveguides.


\begin{center}\textbf{Concept}\end{center}

\begin{figure*}[htbp]
\includegraphics[width = 15cm]{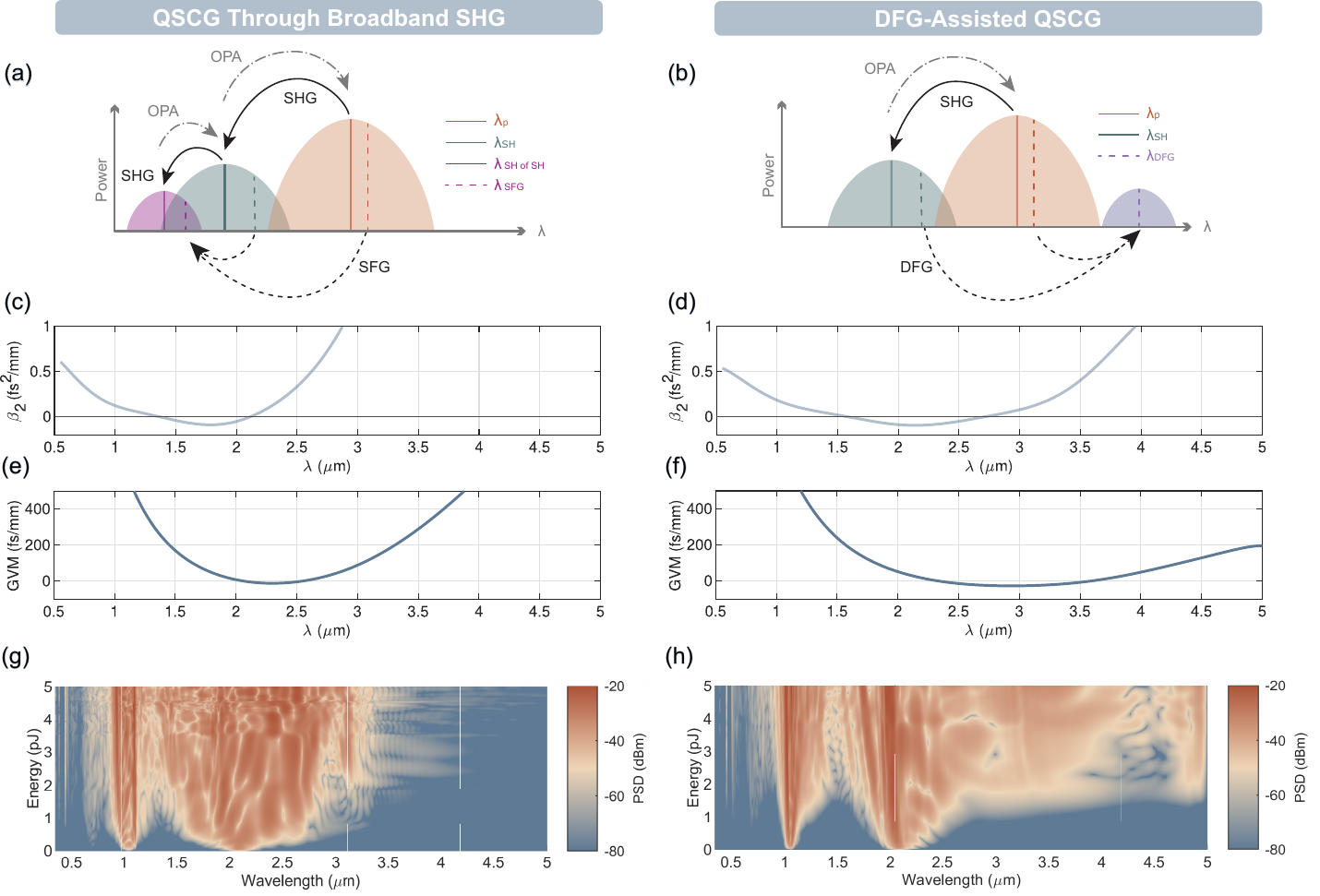}
\caption{QSCG design for efficient broadband SHG (left column) and DFG (right column). \textbf{(a)} Schematic of the spectral broadening mechanism for a quadratic supercontinuum generation (QSCG) waveguide designed for efficient saturated second harmonic generation (SHG) and sum-frequency-generation (SFG). The saturation and back-conversion is enabled by optical parametric amplification (OPA). \textbf{(b)} Schematic of the spectral broadening mechanism for a QSCG waveguide designed for difference frequency generation (DFG) after some initial spectral broadening through saturated SHG. Note that intra-pulse DFG can also occur between the spectral components of the input comb, where the resulting mid-IR comb would have zero $f_{ceo}$. \textbf{(c)} Group velocity dispersion (GVD) for the SHG waveguide. \textbf{(d)} GVD for the DFG waveguide. Compared to the GVD for the SHG waveguide, the zero-crossings are further separated in wavelength. \textbf{(e)} Group velocity mismatch (GVM) between each wavelength and its corresponding SH wavelength for the SHG waveguide. The GVM is near zero at the pump wavelength (2090 nm). \textbf{(f)} GVM between each wavelength and its corresponding SH wavelength for the DFG waveguide. \textbf{(g)},\textbf{(h)} Nonlinear propagation simulation based on the single envelope equation, for the SHG waveguide and DFG waveguide, respectively.}\label{fig1}
\end{figure*}

The main QSCG mechanisms of our work are saturated SHG\,\cite{Marc:23} and intra-pulse DFG. The dispersion engineering strategies for these processes are shown in Fig. \ref{fig1}. In the more conventional saturated SHG method, Fig. \ref{fig1}(a), the desired dispersion profile includes near-zero GVD and GVM between the input and SH wavelengths. With such a dispersion profile, efficient SCG can occur around the input wavelength and its SH through back-and-forth conversion as well as sum-frequency-generation (SFG). The saturation and back-conversion is enabled by optical parametric amplification (OPA). Such an SCG process is robust to QPM variations as the low dispersion provides a broad phase-matching bandwidth for the SHG. However, extension on the long wavelength quickly drops, as the phase matching and dispersion profile are not explicitly designed to support further extension of the spectrum. 

In contrast, one can design the dispersion profile to sacrifice the efficiency of the ideal saturated SHG process to favor DFG processes that lead to the generation of long-wavelength content. As shown in Fig. \ref{fig1}(b), the input spectrum is first doubled through phase-matched (or nearly phase-matched) SHG, followed by subsequent cascaded SHG and SCG through saturated $\chi^{(2)}$ processes. Provided both phase- and group-velocity matching, the spectrally broadened pump and SH can lead to broadband DFG processes to further broaden the spectrum on the long side. Since the DFG frequency is the difference of the higher input frequencies within the pump and SH bandwidths, it naturally produces a longer wavelength ($\lambda_{DFG}$) than the input wavelengths ($\lambda_1$ and $\lambda_2$). Interestingly, for such DFG processes, it is desirable to have far separated zero GVD wavelengths while still maintaining a relatively low GVD around the pump wavelength for the initial pump and SH broadening. Hence, to optimize for spectral broadening into the mid-IR through DFG-assisted QSCG, dispersion parameters and phase-matched poling periods for simultaneous SHG and DFG phase matching should be carefully selected to generate adequate DFG input wavelength pairs, while enabling subsequent broadband DFG phase matching.

We first numerically evaluate the effectiveness of the QSCG design method described above by comparing the dispersion profiles of two TFLN waveguides, at a pump wavelength of 2090 nm. To only highlight the contribution of dispersion and phase matching bandwidth to the spectral broadening processes, both material absorption and third-order nonlinear effects are neglected. Although the dispersion curves are obtained from realistic TFLN waveguides (with Lumerical MODE simulations), their exact geometry parameters are irrelevant for this comparison. The SHG waveguide (Fig. \ref{fig1} left column) is designed to have both near-zero GVD and GVM at the pump wavelength, as shown in Fig. \ref{fig1}(c) and Fig. \ref{fig1}(e), respectively. The DFG waveguide (Fig. \ref{fig1} right column) has non-zero GVD and GVM at the pump wavelength, but has further separated zero GVD wavelengths (at around 1.6 \textmu m and 2.7 \textmu m, Fig. \ref{fig1}(d)) compared to the SHG waveguide. As it can be seen, the GVM curve for the DFG waveguide is overall flatter at longer wavelengths, indicating a wider phase-matching bandwidth critical for wavelength generation above 4 \textmu m, but at the cost of higher dispersion values at the pump.

\begin{figure*}
\includegraphics[width = 15cm]{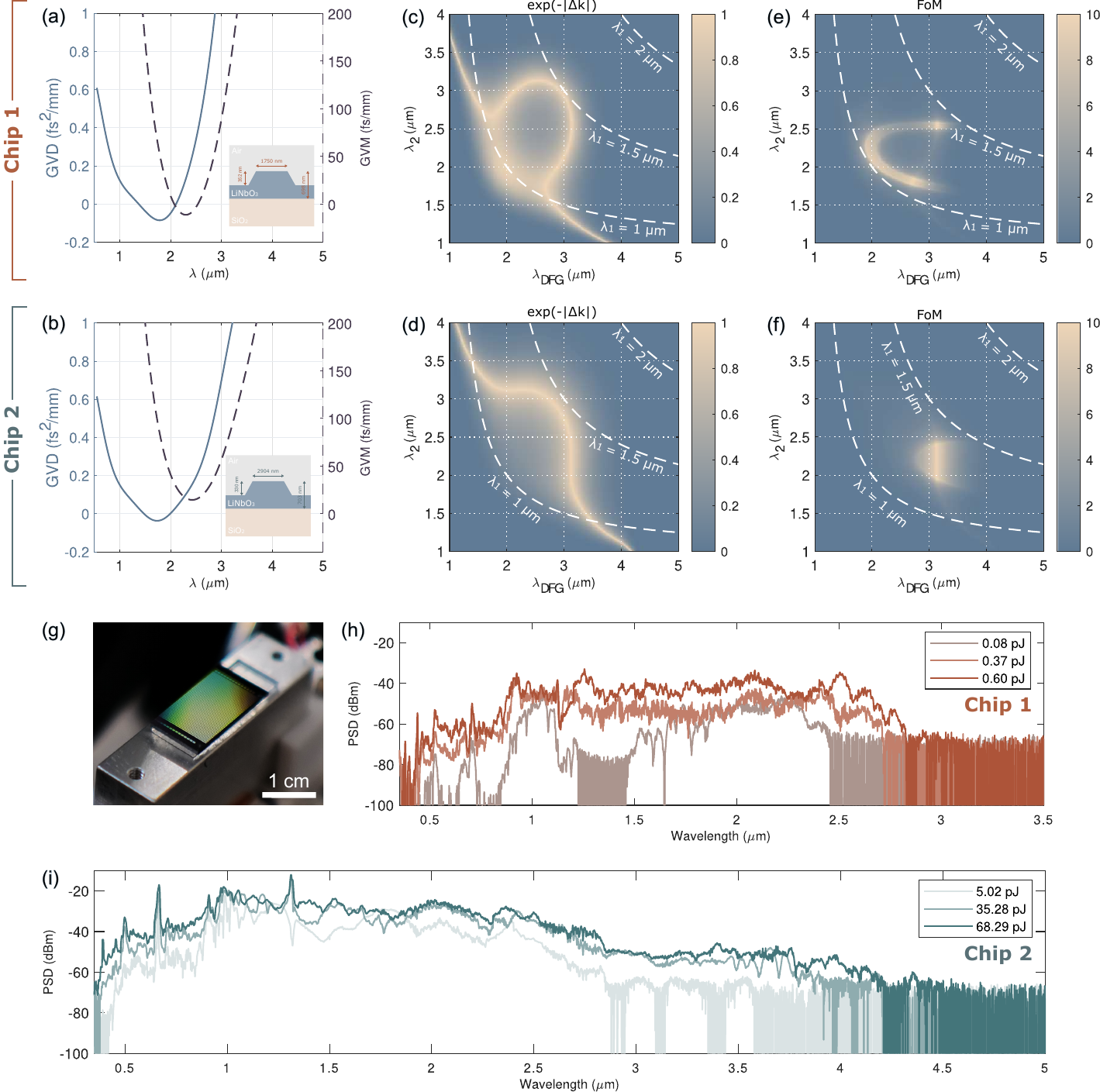}
\caption{TFLN on silica QSCG designs and experimental results. \textbf{(a)} Group velocity dispersion (GVD, blue) and group velocity mismatch (GVM, dashed) for Chip 1, designed for efficient QSCG through SHG. Inset: waveguide geometry for Chip 1. \textbf{(b)} GVD (blue) and GVM (dashed) for Chip 2, designed for mid-IR QSCG through DFG. Inset: waveguide geometry for Chip 2. \textbf{(c)} DFG phase-matching profile for Chip 1. $\lambda_1$ (overlaid contour in dashed while lines) and $\lambda_2$ (plotted on y-axis) denotes the two input wavelengths for generating output DFG wavelengths, $\lambda_{DFG}$ (plotted on x-axis). The color map is exp(-$|\Delta$k$|$), where $\Delta$k is the DFG phase-mismatch in rad/mm for different $\lambda_2$ and $\lambda_{DFG}$ pairs. \textbf{(d)} Same phase-matching profile plot as \textbf{c} for Chip 2. \textbf{(e)} Chip 1 DFG figure of merit (FoM) representing the product between walk-off length and DFG phase-matching (details see Supplementary). \textbf{(f)} Chip 2 DFG FoM. 
\textbf{(g)} Photo of Chip 2 in the experimental setup. 
\textbf{(h)} Experimental saturated-SHG QSCG spectra (Chip 1). 
\textbf{(i)} Experimental DFG-assisted QSCG spectra (Chip 2).} \label{fig2}
\end{figure*}

We next proceed to nonlinear propagation simulations based on the single envelope equation (see Supplementary Section 1) for the two waveguides \cite{Conforti2011}. To ensure a fair comparison, we fixed the effective nonlinear coefficient ($X_0$) for both simulations, which is calculated from the pump and SH fundamental TE mode field overlap for the SHG waveguide geometry and a $\chi^{(2)}$ strength of 36.16\,pm/V. The perfectly phase-matched poling period for each dispersion profile is used, both with a periodically-poled lithium niobate (PPLN) length of 10.8 mm. As shown in Fig. \ref{fig1}(g), the SHG waveguide SCG quickly broadens to a connected spectra between the pump and SH at around 1 pJ of on-chip pulse energy. However, there is negligible frequency content above 3.3\,\textmu m even as the pump power is further increased. In contrast, as shown in Fig. \ref{fig1}(h), the DFG waveguide requires more pulse energy for the initial broadening, but the phase-matched and group-velocity-matched DFG processes quickly extend the spectrum toward 5 \textmu m. These results highlight how dispersion and group-velocity engineering determine whether QSCG favors efficient SH-based broadening or long wavelength extension, guiding subsequent experimental designs presented in the following sections.


\begin{center}\textbf{Experimental DFG-based QSCG}\end{center}

Next, we aim to verify our findings by comparing the saturated SHG QSCG (Fig. \ref{fig1}, left) to the DFG-assisted QSCG (Fig. \ref{fig1}, right) experimentally. 
For that purpose, we design two different chips using the 700-nm-thick TFLN-on-silica platform. The first chip, denoted "Chip 1", is designed for zero GVD and GVM between pump and SH to enable efficient QSCG through saturated SHG. Its dispersion curves and geometry parameters for Chip 1 are shown in Fig. \ref{fig2}(a). Chip 1 dispersion profile exactly matches the one discussed in our theoretical analysis (Fig. \ref{fig1}(c), (e)). 
The second chip, or "Chip 2", is designed for DFG-assisted QSCG. Since the dispersion presented for the DFG-assisted SCG in Fig. \ref{fig1} is not directly achievable with the same TFLN-on-silica wafer, we chose Chip 2 geometry to produce flatter GVD and GVM curves compared to Chip 1, as shown in Fig. \ref{fig2}(b), hence enabling DFG-assisted QSCG. 

To better quantify the efficiency of the DFG process, we introduce a DFG figure of merit (FoM). It consists of the product between walk-off length and DFG phase-matching. The two DFG input wavelengths are denoted $\lambda_1$ and $\lambda_2$, and the resulting output DFG wavelength is denoted $\lambda_{DFG}$. 
The DFG phase-mismatch ($\Delta k$) for Chip 1 and Chip 2 are plotted in Fig. \ref{fig2}(c)-(d) as a 2D colormap. Specifically, we plot $\exp(-|\Delta k|)$ to ease the reading; the higher the value, the more phase-matched the specific DFG process is. More details are given in Supplementary Section 2. The DFG FoM is then found by multiplying the presented phase matching metric by the corresponding walk-off length between $\lambda_1$ and $\lambda_2$. The result for Chip 1 (Chip 2) is plotted in Fig. \ref{fig2}(e) (Fig. \ref{fig2}(f)). 
The comparison between the two clearly highlights that, while the design for Chip 1 favors shorter wavelengths, Chip 2 primarily favors wavelengths above 3 \textmu m through DFG. 

The fabrication steps are discussed in the Supplementary Section 3. A photograph of fabricated Chip 2 is shown in Fig. \ref{fig2}(g). We note that Chip 1, presenting a zero-GVM at pump, has a longer PPLN length (10.8 mm) than Chip 2 (6.5 mm).  
Once fabricated, we pump both devices with transform-limited 45\,fs pulses centered at $2090$\,nm, produced by a 250\,MHz repetition rate optical parametric oscillator. The two experimental quadratic supercontinuum are plotted in Fig. \ref{fig2}(h), (i) for different pump powers. While the saturated SHG device (Chip 1) leads to highly efficient SCG between the pump and SH signal due to the low GVD and GVM, and has a relatively long PPLN length, increasing the pump power does not further broaden the spectra above 3 \textmu m. This is consistent with our numerical simulation previously shown in Fig. \ref{fig1}(g), verifying the trade-off between efficient QSCG through saturated SHG and long wavelength formation.
In contrast, for the DFG-assisted QSCG chip (Chip 2), we observe the formation of spectral content past the SiO$_2$ absorption band and up to 4.2 \textmu m as the pump power is further increased, despite a lower cascaded SHG efficiency. To the best of our knowledge, this is the longest wavelength generated through QSCG using the TFLN-on-silica platform. 
Although these results are not a direct side-to-side comparison due to non-negligible differences in coupling loss and PPLN lengths between the two chips, they clearly show that different design strategies should be employed when targeting either efficient spectral broadening or mid-IR wavelength generation. The results from Chip 2 highlight the merits of the DFG approach for generating frequency content in the mid-IR. 

\begin{center}\textbf{DFG-assisted QSCG across the lithium niobate transparency window}\end{center}

\begin{figure*}
\includegraphics[width = 15cm]{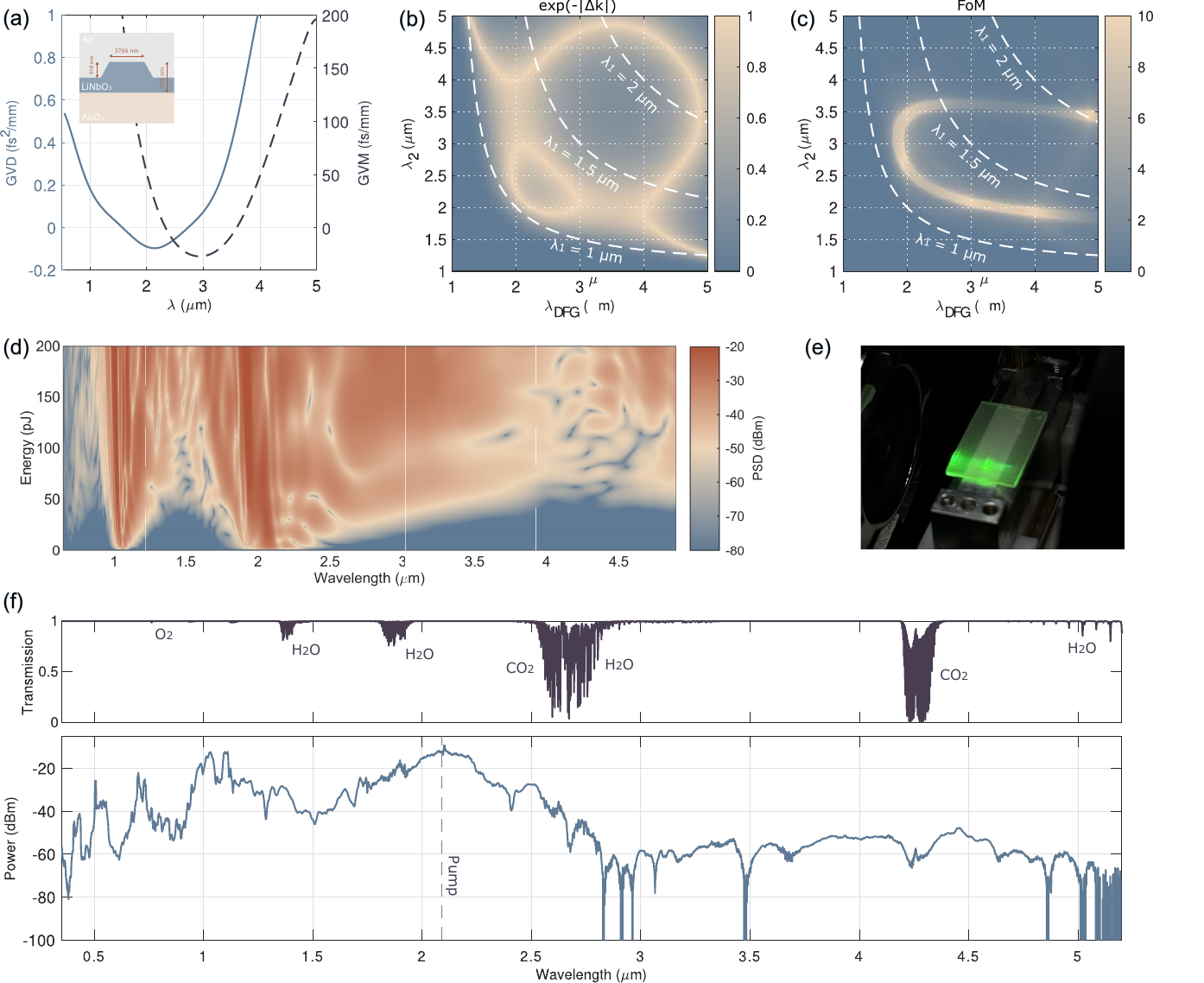}
\caption{TFLN on sapphire QSCG. \textbf{(a)} Group velocity dispersion (GVD, blue) and group velocity mismatch (GVM, dashed) curves for Chip 3, designed for efficient mid-IR QSCG through SHG. Inset: waveguide geometry parameters for Chip 3. \textbf{(b)} DFG phase-matching profile for Chip 3, $\lambda_1$ (overlaid contour in dashed while lines) and $\lambda_2$ (plotted on y-axis) denote the two input wavelengths for generating output DFG wavelengths, $\lambda_{DFG}$ (plotted on x-axis). The color map is exp(-$|\Delta$k$|$), where $\Delta$k is the DFG phase-mismatch in rad/mm for different $\lambda_2$ and $\lambda_{DFG}$ pairs. \textbf{(c)} Chip 3 DFG figure of merit (FoM) representing the product between walk-off length and DFG phase-matching (details see Supplementary). 
\textbf{(d)} Nonlinear propagation simulations based on the single envelope equation for Chip 3, with $\chi_{eff}^{(2)}$ divided by 4 (compared to Chip 1 and Chip 2) to account for lower poling quality. 
\textbf{(e)} Photo of Chip 3 on the experimental setup. 
\textbf{(f)} Top: combined transmission spectra over 50 cm (corresponding to the free-space optical path between the waveguide output facet and free-space to fiber collimator) for atmospheric CO$_2$, N$_2$, O$_2$, and H$_2$O\,\cite{HITRAN}. Bottom: experimental QSCG spectra for Chip 3, with 177 pJ of on chip pump pulse energy. }\label{fig3}
\end{figure*}

Aside from dispersion engineering and DFG phase-matching strategies, replacing the silica substrate with sapphire is an additional design consideration to further extend the bandwidth of DFG-assisted QSCG in the mid-IR. To show it, we design and fabricate a third device, Chip 3, using the 990-nm-thick TFLN-on-sapphire platform.
We chose its geometry such that its dispersion parameters (Fig. \ref{fig3}(a)) exactly coincide with those used in our theoretical analysis of the DFG-assisted QSCG (Fig. \ref{fig1}, right column).
As for the first two devices, we calculate the theoretical DFG phase matching profile and FoM for Chip 3. The results are shown in Fig. \ref{fig3}(b) and Fig. \ref{fig3}(c), respectively.
Compared to both other chips, Chip 3 supports a wider DFG phase-matching bandwidth. In addition, its FoM plot indicates efficient DFG generation from pump wavelength up to 5 \textmu m (see also Supplementary Section 6).
As an additional verification, we simulate the propagation in Chip 3, anticipating the experimental parameters. In particular, we reduce the nonlinear efficiency by a factor of four to account for a poorer poling quality on the TFLN on Sapphire platform (see Supplementary Section 3). Despite the lowered efficiency, the results of our simulation plotted in Fig. \ref{fig3}(d) show that, for pump power as low as 150\,pJ, the DFG-assisted QSCG spans the entire lithium niobate transparency window. 

To verify these theoretical and numerical predictions, we fabricated Chip 3 following a similar procedure to that of Chips 1 and 2, with a PPLN length of 7 mm. Compared to Chip 2, Chip 3 has a larger pump GVD (-5 fs$^2$/mm) and GVM (32 fs/mm) between the pump and SH, and a longer (7\,mm) poling section. A photograph of the chip is shown in Fig. \ref{fig3}(e). Similarly to Chips 1 and 2, we pump Chip 3 with 45\,fs-long pulses centered at 2090\,nm. 
The experimental QSCG, for 177 pJ of estimated on-chip pump pulse energy\cite{gray2025,Ledezma:22}, is plotted in Fig. \ref{fig3}(f) (bottom). It spans 3.8 octaves, essentially covering the entire transparency window of lithium niobate, in qualitatively good agreement with our numerical simulations (Fig. \ref{fig3}(e)). Some of its spectral features can be attributed to atmospheric absorption over the 50 cm of propagation (free-space optical path between the waveguide output facet and free-space to fiber collimator). For instance, the CO$_2$ absorption dip around 4.3-4.3 \textmu m is clearly visible on the QSCG trace, limiting power around that wavelength region.


\begin{center}\textbf{Discussion}\end{center}

In conclusion, we have demonstrated purely quadratic frequency comb generation in the mid-IR through simultaneous SHG, SCG, and DFG processes in dispersion-engineered and quasi-phase-matched waveguides in TFLN on silica and sapphire. Our results show that balancing low GVD and GVM for the pump and SH signals with broadband DFG phase matching is crucial for extending the mid-IR spectrum efficiency. Chip 1, designed with near-zero GVD and GVM, achieved more than two octaves QSCG with sub-pJ pump pulse energy, which is a few orders of magnitude less energy compared to other comparable spectra achieved in single-pass devices (see Supplementary Table S1). To our knowledge, Chip 2 represents the highest-performing QSCG into the mid-IR on a silica substrate platform to date, enabled by optimized dispersion and a relatively flat group-index profile, reaching beyond 4.2 \textmu m despite silica’s intrinsic absorption. By implementing the same principles on a TFLN-on-sapphire platform (Chip 3), we extended mid-IR generation beyond 5\,\textmu m to fill the entire transparency window of lithium niobate. 
While the substrate absorption, although depending on the mode overlap, may not fundamentally limit spectral extension if dispersion and phase-matching conditions are properly engineered, the TFLN-on-sapphire platform can enable QSCG across the entire transparency window of lithium niobate.
The versatility of QSCG on TFLN currently allows for designs that prioritize either efficiency or spectral coverage.

Several improvements to our work are foreseen. 
First, the poling quality for TFLN-on-sapphire can be improved, and the coupling losses decreased, to lower the pump power requirements.  
Additionally, using multi-parameter optimization processes, one can potentially find geometries that simultaneously achieve both low GVD and low GVM at the pump wavelength, as well as a broad DFG phase matching bandwidth into the mid-IR. Such a design would simultaneously support mid-IR generation through DFG-assisted QSCG, as well as efficient initial broadening through SHG. Alternatively, cascading two QSCG devices can potentially achieve similar enhanced performance. Finally, active control of QPM along the waveguide length to dynamically tune phase-matching conditions can also be investigated. 

\begin{center}\textbf{Acknowledgments}\end{center}

The device nanofabrication was performed at the Kavli Nanoscience Institute (KNI) at Caltech and the UCLA NanoLab. The authors gratefully acknowledge support from DARPA award D23AP00158, ARO grant no.W911NF-23-1-0048, NSF grant no. 2408297, 1918549, AFOSR award FA9550-23-1-0755, the Center for Sensing to Intelligence at Caltech, the Alfred P. Sloan Foundation, and NASA/JPL. N.E. acknowledges support from the Belgian American Educational Foundation (B.A.E.F.) and the European Union's Horizon Europe research and innovation programme under the Marie Skłodowska-Curie Grant Agreement No. 101103780.

\hfill

\noindent \textbf{Competing Interests:} R.S., R.M.G., and A.M. are inventors on a U.S. provisional patent application filed by the California Institute of Technology (application number 18/805,408). R.S. and A.M. are involved in developing photonic integrated nonlinear circuits at PINC Technologies Inc. R.S. and A.M. have an equity interest in PINC Technologies Inc.

\hfill

\noindent \textbf{Data Availability:} The data used for generation of the figures within this manuscript and other findings of this study are available upon request from the corresponding author.


\bibliographystyle{apsrev4-2}  
\input{Main.bbl}

\end{document}

%% file: Main.bbl
%

%% file: Main.bbl
\begin{thebibliography}{49}%
\makeatletter
\providecommand \@ifxundefined [1]{%
 \@ifx{#1\undefined}
}%
\providecommand \@ifnum [1]{%
 \ifnum #1\expandafter \@firstoftwo
 \else \expandafter \@secondoftwo
 \fi
}%
\providecommand \@ifx [1]{%
 \ifx #1\expandafter \@firstoftwo
 \else \expandafter \@secondoftwo
 \fi
}%
\providecommand \natexlab [1]{#1}%
\providecommand \enquote  [1]{``#1''}%
\providecommand \bibnamefont  [1]{#1}%
\providecommand \bibfnamefont [1]{#1}%
\providecommand \citenamefont [1]{#1}%
\providecommand \href@noop [0]{\@secondoftwo}%
\providecommand \href [0]{\begingroup \@sanitize@url \@href}%
\providecommand \@href[1]{\@@startlink{#1}\@@href}%
\providecommand \@@href[1]{\endgroup#1\@@endlink}%
\providecommand \@sanitize@url [0]{\catcode `\\12\catcode `\$12\catcode `\&12\catcode `\#12\catcode `\^12\catcode `\_12\catcode `\%12\relax}%
\providecommand \@@startlink[1]{}%
\providecommand \@@endlink[0]{}%
\providecommand \url  [0]{\begingroup\@sanitize@url \@url }%
\providecommand \@url [1]{\endgroup\@href {#1}{\urlprefix }}%
\providecommand \urlprefix  [0]{URL }%
\providecommand \Eprint [0]{\href }%
\providecommand \doibase [0]{https://doi.org/}%
\providecommand \selectlanguage [0]{\@gobble}%
\providecommand \bibinfo  [0]{\@secondoftwo}%
\providecommand \bibfield  [0]{\@secondoftwo}%
\providecommand \translation [1]{[#1]}%
\providecommand \BibitemOpen [0]{}%
\providecommand \bibitemStop [0]{}%
\providecommand \bibitemNoStop [0]{.\EOS\space}%
\providecommand \EOS [0]{\spacefactor3000\relax}%
\providecommand \BibitemShut  [1]{\csname bibitem#1\endcsname}%
\let\auto@bib@innerbib\@empty
\bibitem [{\citenamefont {Kippenberg}\ \emph {et~al.}(2018)\citenamefont {Kippenberg}, \citenamefont {Gaeta}, \citenamefont {Lipson},\ and\ \citenamefont {Gorodetsky}}]{kippenberg_dissipative_2018}%
  \BibitemOpen
  \bibfield  {author} {\bibinfo {author} {\bibfnamefont {T.~J.}\ \bibnamefont {Kippenberg}}, \bibinfo {author} {\bibfnamefont {A.~L.}\ \bibnamefont {Gaeta}}, \bibinfo {author} {\bibfnamefont {M.}~\bibnamefont {Lipson}},\ and\ \bibinfo {author} {\bibfnamefont {M.~L.}\ \bibnamefont {Gorodetsky}},\ }\bibfield  {journal} {\bibinfo  {journal} {Science}\ }\textbf {\bibinfo {volume} {361}},\ \href {https://doi.org/10.1126/science.aan8083} {10.1126/science.aan8083} (\bibinfo {year} {2018}),\ \bibinfo {note} {publisher: American Association for the Advancement of Science Section: Review}\BibitemShut {NoStop}%
\bibitem [{\citenamefont {Pasquazi}\ \emph {et~al.}(2018)\citenamefont {Pasquazi}, \citenamefont {Peccianti}, \citenamefont {Razzari}, \citenamefont {Moss}, \citenamefont {Coen}, \citenamefont {Erkintalo}, \citenamefont {Chembo}, \citenamefont {Hansson}, \citenamefont {Wabnitz}, \citenamefont {Del’Haye}, \citenamefont {Xue}, \citenamefont {Weiner},\ and\ \citenamefont {Morandotti}}]{pasquazi_micro-combs_2018}%
  \BibitemOpen
  \bibfield  {author} {\bibinfo {author} {\bibfnamefont {A.}~\bibnamefont {Pasquazi}}, \bibinfo {author} {\bibfnamefont {M.}~\bibnamefont {Peccianti}}, \bibinfo {author} {\bibfnamefont {L.}~\bibnamefont {Razzari}}, \bibinfo {author} {\bibfnamefont {D.~J.}\ \bibnamefont {Moss}}, \bibinfo {author} {\bibfnamefont {S.}~\bibnamefont {Coen}}, \bibinfo {author} {\bibfnamefont {M.}~\bibnamefont {Erkintalo}}, \bibinfo {author} {\bibfnamefont {Y.~K.}\ \bibnamefont {Chembo}}, \bibinfo {author} {\bibfnamefont {T.}~\bibnamefont {Hansson}}, \bibinfo {author} {\bibfnamefont {S.}~\bibnamefont {Wabnitz}}, \bibinfo {author} {\bibfnamefont {P.}~\bibnamefont {Del’Haye}}, \bibinfo {author} {\bibfnamefont {X.}~\bibnamefont {Xue}}, \bibinfo {author} {\bibfnamefont {A.~M.}\ \bibnamefont {Weiner}},\ and\ \bibinfo {author} {\bibfnamefont {R.}~\bibnamefont {Morandotti}},\ }\href {https://doi.org/10.1016/j.physrep.2017.08.004} {\bibfield  {journal} {\bibinfo  {journal} {Physics Reports}\ }\bibinfo {series} {Micro-combs: {A} novel
  generation of optical sources},\ \textbf {\bibinfo {volume} {729}},\ \bibinfo {pages} {1} (\bibinfo {year} {2018})}\BibitemShut {NoStop}%
\bibitem [{\citenamefont {Diddams}\ \emph {et~al.}(2020)\citenamefont {Diddams}, \citenamefont {Vahala},\ and\ \citenamefont {Udem}}]{diddams_optical_2020}%
  \BibitemOpen
  \bibfield  {author} {\bibinfo {author} {\bibfnamefont {S.~A.}\ \bibnamefont {Diddams}}, \bibinfo {author} {\bibfnamefont {K.}~\bibnamefont {Vahala}},\ and\ \bibinfo {author} {\bibfnamefont {T.}~\bibnamefont {Udem}},\ }\href {https://doi.org/10.1126/science.aay3676} {\bibfield  {journal} {\bibinfo  {journal} {Science}\ }\textbf {\bibinfo {volume} {369}},\ \bibinfo {pages} {eaay3676} (\bibinfo {year} {2020})},\ \bibinfo {note} {publisher: American Association for the Advancement of Science}\BibitemShut {NoStop}%
\bibitem [{\citenamefont {Davenport}\ \emph {et~al.}(2018)\citenamefont {Davenport}, \citenamefont {Liu},\ and\ \citenamefont {Bowers}}]{davenport_integrated_2018}%
  \BibitemOpen
  \bibfield  {author} {\bibinfo {author} {\bibfnamefont {M.~L.}\ \bibnamefont {Davenport}}, \bibinfo {author} {\bibfnamefont {S.}~\bibnamefont {Liu}},\ and\ \bibinfo {author} {\bibfnamefont {J.~E.}\ \bibnamefont {Bowers}},\ }\href {https://doi.org/10.1364/PRJ.6.000468} {\bibfield  {journal} {\bibinfo  {journal} {Photonics Research}\ }\textbf {\bibinfo {volume} {6}},\ \bibinfo {pages} {468} (\bibinfo {year} {2018})},\ \bibinfo {note} {publisher: Optica Publishing Group}\BibitemShut {NoStop}%
\bibitem [{\citenamefont {Gaeta}\ \emph {et~al.}(2019{\natexlab{a}})\citenamefont {Gaeta}, \citenamefont {Lipson},\ and\ \citenamefont {Kippenberg}}]{gaeta_photonic-chip-based_2019}%
  \BibitemOpen
  \bibfield  {author} {\bibinfo {author} {\bibfnamefont {A.~L.}\ \bibnamefont {Gaeta}}, \bibinfo {author} {\bibfnamefont {M.}~\bibnamefont {Lipson}},\ and\ \bibinfo {author} {\bibfnamefont {T.~J.}\ \bibnamefont {Kippenberg}},\ }\href {https://doi.org/10.1038/s41566-019-0358-x} {\bibfield  {journal} {\bibinfo  {journal} {Nature Photonics}\ }\textbf {\bibinfo {volume} {13}},\ \bibinfo {pages} {158} (\bibinfo {year} {2019}{\natexlab{a}})},\ \bibinfo {note} {publisher: Nature Publishing Group}\BibitemShut {NoStop}%
\bibitem [{\citenamefont {Van~Gasse}\ \emph {et~al.}(2019)\citenamefont {Van~Gasse}, \citenamefont {Uvin}, \citenamefont {Moskalenko}, \citenamefont {Latkowski}, \citenamefont {Roelkens}, \citenamefont {Bente},\ and\ \citenamefont {kuyken}}]{van_gasse_recent_2019}%
  \BibitemOpen
  \bibfield  {author} {\bibinfo {author} {\bibfnamefont {K.}~\bibnamefont {Van~Gasse}}, \bibinfo {author} {\bibfnamefont {S.}~\bibnamefont {Uvin}}, \bibinfo {author} {\bibfnamefont {V.}~\bibnamefont {Moskalenko}}, \bibinfo {author} {\bibfnamefont {S.}~\bibnamefont {Latkowski}}, \bibinfo {author} {\bibfnamefont {G.}~\bibnamefont {Roelkens}}, \bibinfo {author} {\bibfnamefont {E.}~\bibnamefont {Bente}},\ and\ \bibinfo {author} {\bibfnamefont {B.}~\bibnamefont {kuyken}},\ }\href {https://doi.org/10.1109/LPT.2019.2945973} {\bibfield  {journal} {\bibinfo  {journal} {IEEE Photonics Technology Letters}\ }\textbf {\bibinfo {volume} {31}},\ \bibinfo {pages} {1870} (\bibinfo {year} {2019})}\BibitemShut {NoStop}%
\bibitem [{\citenamefont {Chang}\ \emph {et~al.}(2022)\citenamefont {Chang}, \citenamefont {Liu},\ and\ \citenamefont {Bowers}}]{chang_integrated_2022}%
  \BibitemOpen
  \bibfield  {author} {\bibinfo {author} {\bibfnamefont {L.}~\bibnamefont {Chang}}, \bibinfo {author} {\bibfnamefont {S.}~\bibnamefont {Liu}},\ and\ \bibinfo {author} {\bibfnamefont {J.~E.}\ \bibnamefont {Bowers}},\ }\href {https://doi.org/10.1038/s41566-021-00945-1} {\bibfield  {journal} {\bibinfo  {journal} {Nature Photonics}\ }\textbf {\bibinfo {volume} {16}},\ \bibinfo {pages} {95} (\bibinfo {year} {2022})},\ \bibinfo {note} {publisher: Nature Publishing Group}\BibitemShut {NoStop}%
\bibitem [{\citenamefont {Meng}\ \emph {et~al.}(2022)\citenamefont {Meng}, \citenamefont {Singleton}, \citenamefont {Hillbrand}, \citenamefont {Franckié}, \citenamefont {Beck},\ and\ \citenamefont {Faist}}]{meng_dissipative_2022}%
  \BibitemOpen
  \bibfield  {author} {\bibinfo {author} {\bibfnamefont {B.}~\bibnamefont {Meng}}, \bibinfo {author} {\bibfnamefont {M.}~\bibnamefont {Singleton}}, \bibinfo {author} {\bibfnamefont {J.}~\bibnamefont {Hillbrand}}, \bibinfo {author} {\bibfnamefont {M.}~\bibnamefont {Franckié}}, \bibinfo {author} {\bibfnamefont {M.}~\bibnamefont {Beck}},\ and\ \bibinfo {author} {\bibfnamefont {J.}~\bibnamefont {Faist}},\ }\href {https://doi.org/10.1038/s41566-021-00927-3} {\bibfield  {journal} {\bibinfo  {journal} {Nature Photonics}\ }\textbf {\bibinfo {volume} {16}},\ \bibinfo {pages} {142} (\bibinfo {year} {2022})},\ \bibinfo {note} {publisher: Nature Publishing Group}\BibitemShut {NoStop}%
\bibitem [{\citenamefont {Kazakov}\ \emph {et~al.}(2025)\citenamefont {Kazakov}, \citenamefont {Letsou}, \citenamefont {Piccardo}, \citenamefont {Columbo}, \citenamefont {Brambilla}, \citenamefont {Prati}, \citenamefont {Dal~Cin}, \citenamefont {Beiser}, \citenamefont {Opačak}, \citenamefont {Ratra}, \citenamefont {Pushkarsky}, \citenamefont {Caffey}, \citenamefont {Day}, \citenamefont {Lugiato}, \citenamefont {Schwarz},\ and\ \citenamefont {Capasso}}]{kazakov_driven_2025}%
  \BibitemOpen
  \bibfield  {author} {\bibinfo {author} {\bibfnamefont {D.}~\bibnamefont {Kazakov}}, \bibinfo {author} {\bibfnamefont {T.~P.}\ \bibnamefont {Letsou}}, \bibinfo {author} {\bibfnamefont {M.}~\bibnamefont {Piccardo}}, \bibinfo {author} {\bibfnamefont {L.~L.}\ \bibnamefont {Columbo}}, \bibinfo {author} {\bibfnamefont {M.}~\bibnamefont {Brambilla}}, \bibinfo {author} {\bibfnamefont {F.}~\bibnamefont {Prati}}, \bibinfo {author} {\bibfnamefont {S.}~\bibnamefont {Dal~Cin}}, \bibinfo {author} {\bibfnamefont {M.}~\bibnamefont {Beiser}}, \bibinfo {author} {\bibfnamefont {N.}~\bibnamefont {Opačak}}, \bibinfo {author} {\bibfnamefont {P.}~\bibnamefont {Ratra}}, \bibinfo {author} {\bibfnamefont {M.}~\bibnamefont {Pushkarsky}}, \bibinfo {author} {\bibfnamefont {D.}~\bibnamefont {Caffey}}, \bibinfo {author} {\bibfnamefont {T.}~\bibnamefont {Day}}, \bibinfo {author} {\bibfnamefont {L.~A.}\ \bibnamefont {Lugiato}}, \bibinfo {author} {\bibfnamefont {B.}~\bibnamefont {Schwarz}},\ and\ \bibinfo {author} {\bibfnamefont
  {F.}~\bibnamefont {Capasso}},\ }\href {https://doi.org/10.1038/s41586-025-08853-y} {\bibfield  {journal} {\bibinfo  {journal} {Nature}\ }\textbf {\bibinfo {volume} {641}},\ \bibinfo {pages} {83} (\bibinfo {year} {2025})},\ \bibinfo {note} {publisher: Nature Publishing Group}\BibitemShut {NoStop}%
\bibitem [{\citenamefont {Pfeiffer}\ \emph {et~al.}(2017)\citenamefont {Pfeiffer}, \citenamefont {Herkommer}, \citenamefont {Liu}, \citenamefont {Guo}, \citenamefont {Karpov}, \citenamefont {Lucas}, \citenamefont {Zervas},\ and\ \citenamefont {Kippenberg}}]{Pfeiffer:17}%
  \BibitemOpen
  \bibfield  {author} {\bibinfo {author} {\bibfnamefont {M.~H.~P.}\ \bibnamefont {Pfeiffer}}, \bibinfo {author} {\bibfnamefont {C.}~\bibnamefont {Herkommer}}, \bibinfo {author} {\bibfnamefont {J.}~\bibnamefont {Liu}}, \bibinfo {author} {\bibfnamefont {H.}~\bibnamefont {Guo}}, \bibinfo {author} {\bibfnamefont {M.}~\bibnamefont {Karpov}}, \bibinfo {author} {\bibfnamefont {E.}~\bibnamefont {Lucas}}, \bibinfo {author} {\bibfnamefont {M.}~\bibnamefont {Zervas}},\ and\ \bibinfo {author} {\bibfnamefont {T.~J.}\ \bibnamefont {Kippenberg}},\ }\href {https://doi.org/10.1364/OPTICA.4.000684} {\bibfield  {journal} {\bibinfo  {journal} {Optica}\ }\textbf {\bibinfo {volume} {4}},\ \bibinfo {pages} {684} (\bibinfo {year} {2017})}\BibitemShut {NoStop}%
\bibitem [{\citenamefont {Brasch}\ \emph {et~al.}(2017)\citenamefont {Brasch}, \citenamefont {Lucas}, \citenamefont {Jost}, \citenamefont {Geiselmann},\ and\ \citenamefont {Kippenberg}}]{Brasch2017}%
  \BibitemOpen
  \bibfield  {author} {\bibinfo {author} {\bibfnamefont {V.}~\bibnamefont {Brasch}}, \bibinfo {author} {\bibfnamefont {E.}~\bibnamefont {Lucas}}, \bibinfo {author} {\bibfnamefont {J.~D.}\ \bibnamefont {Jost}}, \bibinfo {author} {\bibfnamefont {M.}~\bibnamefont {Geiselmann}},\ and\ \bibinfo {author} {\bibfnamefont {T.~J.}\ \bibnamefont {Kippenberg}},\ }\href {https://doi.org/10.1038/lsa.2016.202} {\bibfield  {journal} {\bibinfo  {journal} {Light: Science {\&} Applications}\ }\textbf {\bibinfo {volume} {6}},\ \bibinfo {pages} {e16202} (\bibinfo {year} {2017})}\BibitemShut {NoStop}%
\bibitem [{\citenamefont {Gaeta}\ \emph {et~al.}(2019{\natexlab{b}})\citenamefont {Gaeta}, \citenamefont {Lipson},\ and\ \citenamefont {Kippenberg}}]{Gaeta19_Review}%
  \BibitemOpen
  \bibfield  {author} {\bibinfo {author} {\bibfnamefont {A.~L.}\ \bibnamefont {Gaeta}}, \bibinfo {author} {\bibfnamefont {M.}~\bibnamefont {Lipson}},\ and\ \bibinfo {author} {\bibfnamefont {T.~J.}\ \bibnamefont {Kippenberg}},\ }\href {https://doi.org/10.1038/s41566-019-0358-x} {\bibfield  {journal} {\bibinfo  {journal} {Nature Photonics}\ }\textbf {\bibinfo {volume} {13}},\ \bibinfo {pages} {158} (\bibinfo {year} {2019}{\natexlab{b}})}\BibitemShut {NoStop}%
\bibitem [{\citenamefont {Brès}\ \emph {et~al.}(2023)\citenamefont {Brès}, \citenamefont {Torre}, \citenamefont {Grassani}, \citenamefont {Brasch}, \citenamefont {Grillet},\ and\ \citenamefont {Monat}}]{Bres23_Review}%
  \BibitemOpen
  \bibfield  {author} {\bibinfo {author} {\bibfnamefont {C.-S.}\ \bibnamefont {Brès}}, \bibinfo {author} {\bibfnamefont {A.~D.}\ \bibnamefont {Torre}}, \bibinfo {author} {\bibfnamefont {D.}~\bibnamefont {Grassani}}, \bibinfo {author} {\bibfnamefont {V.}~\bibnamefont {Brasch}}, \bibinfo {author} {\bibfnamefont {C.}~\bibnamefont {Grillet}},\ and\ \bibinfo {author} {\bibfnamefont {C.}~\bibnamefont {Monat}},\ }\href {https://doi.org/doi:10.1515/nanoph-2022-0749} {\bibfield  {journal} {\bibinfo  {journal} {Nanophotonics}\ }\textbf {\bibinfo {volume} {12}},\ \bibinfo {pages} {1199} (\bibinfo {year} {2023})}\BibitemShut {NoStop}%
\bibitem [{\citenamefont {Suh}\ \emph {et~al.}(2016)\citenamefont {Suh}, \citenamefont {Yang}, \citenamefont {Yang}, \citenamefont {Yi},\ and\ \citenamefont {Vahala}}]{MGSuh16}%
  \BibitemOpen
  \bibfield  {author} {\bibinfo {author} {\bibfnamefont {M.-G.}\ \bibnamefont {Suh}}, \bibinfo {author} {\bibfnamefont {Q.-F.}\ \bibnamefont {Yang}}, \bibinfo {author} {\bibfnamefont {K.~Y.}\ \bibnamefont {Yang}}, \bibinfo {author} {\bibfnamefont {X.}~\bibnamefont {Yi}},\ and\ \bibinfo {author} {\bibfnamefont {K.~J.}\ \bibnamefont {Vahala}},\ }\href {https://doi.org/10.1126/science.aah6516} {\bibfield  {journal} {\bibinfo  {journal} {Science}\ }\textbf {\bibinfo {volume} {354}},\ \bibinfo {pages} {600} (\bibinfo {year} {2016})}\BibitemShut {NoStop}%
\bibitem [{\citenamefont {Stern}\ \emph {et~al.}(2020)\citenamefont {Stern}, \citenamefont {Stone}, \citenamefont {Kang}, \citenamefont {Cole}, \citenamefont {Suh}, \citenamefont {Fredrick}, \citenamefont {Newman}, \citenamefont {Vahala}, \citenamefont {Kitching}, \citenamefont {Diddams},\ and\ \citenamefont {Papp}}]{Stern:2020}%
  \BibitemOpen
  \bibfield  {author} {\bibinfo {author} {\bibfnamefont {L.}~\bibnamefont {Stern}}, \bibinfo {author} {\bibfnamefont {J.~R.}\ \bibnamefont {Stone}}, \bibinfo {author} {\bibfnamefont {S.}~\bibnamefont {Kang}}, \bibinfo {author} {\bibfnamefont {D.~C.}\ \bibnamefont {Cole}}, \bibinfo {author} {\bibfnamefont {M.-G.}\ \bibnamefont {Suh}}, \bibinfo {author} {\bibfnamefont {C.}~\bibnamefont {Fredrick}}, \bibinfo {author} {\bibfnamefont {Z.}~\bibnamefont {Newman}}, \bibinfo {author} {\bibfnamefont {K.}~\bibnamefont {Vahala}}, \bibinfo {author} {\bibfnamefont {J.}~\bibnamefont {Kitching}}, \bibinfo {author} {\bibfnamefont {S.~A.}\ \bibnamefont {Diddams}},\ and\ \bibinfo {author} {\bibfnamefont {S.~B.}\ \bibnamefont {Papp}},\ }\href {https://doi.org/10.1126/sciadv.aax6230} {\bibfield  {journal} {\bibinfo  {journal} {Science Advances}\ }\textbf {\bibinfo {volume} {6}},\ \bibinfo {pages} {eaax6230} (\bibinfo {year} {2020})}\BibitemShut {NoStop}%
\bibitem [{\citenamefont {Bao}\ \emph {et~al.}(2021)\citenamefont {Bao}, \citenamefont {Yuan}, \citenamefont {Wu}, \citenamefont {Suh}, \citenamefont {Wang}, \citenamefont {Lin},\ and\ \citenamefont {Vahala}}]{Bao2021}%
  \BibitemOpen
  \bibfield  {author} {\bibinfo {author} {\bibfnamefont {C.}~\bibnamefont {Bao}}, \bibinfo {author} {\bibfnamefont {Z.}~\bibnamefont {Yuan}}, \bibinfo {author} {\bibfnamefont {L.}~\bibnamefont {Wu}}, \bibinfo {author} {\bibfnamefont {M.-G.}\ \bibnamefont {Suh}}, \bibinfo {author} {\bibfnamefont {H.}~\bibnamefont {Wang}}, \bibinfo {author} {\bibfnamefont {Q.}~\bibnamefont {Lin}},\ and\ \bibinfo {author} {\bibfnamefont {K.~J.}\ \bibnamefont {Vahala}},\ }\href {https://doi.org/10.1038/s41467-021-26958-6} {\bibfield  {journal} {\bibinfo  {journal} {Nature Communications}\ }\textbf {\bibinfo {volume} {12}},\ \bibinfo {pages} {6573} (\bibinfo {year} {2021})}\BibitemShut {NoStop}%
\bibitem [{\citenamefont {Singh}\ \emph {et~al.}(2020)\citenamefont {Singh}, \citenamefont {Xin}, \citenamefont {Li}, \citenamefont {Vermeulen}, \citenamefont {Ruocco}, \citenamefont {Magden}, \citenamefont {Shtyrkova}, \citenamefont {Ippen}, \citenamefont {Kärtner},\ and\ \citenamefont {Watts}}]{Singh20}%
  \BibitemOpen
  \bibfield  {author} {\bibinfo {author} {\bibfnamefont {N.}~\bibnamefont {Singh}}, \bibinfo {author} {\bibfnamefont {M.}~\bibnamefont {Xin}}, \bibinfo {author} {\bibfnamefont {N.}~\bibnamefont {Li}}, \bibinfo {author} {\bibfnamefont {D.}~\bibnamefont {Vermeulen}}, \bibinfo {author} {\bibfnamefont {A.}~\bibnamefont {Ruocco}}, \bibinfo {author} {\bibfnamefont {E.~S.}\ \bibnamefont {Magden}}, \bibinfo {author} {\bibfnamefont {K.}~\bibnamefont {Shtyrkova}}, \bibinfo {author} {\bibfnamefont {E.}~\bibnamefont {Ippen}}, \bibinfo {author} {\bibfnamefont {F.~X.}\ \bibnamefont {Kärtner}},\ and\ \bibinfo {author} {\bibfnamefont {M.~R.}\ \bibnamefont {Watts}},\ }\href {https://doi.org/https://doi.org/10.1002/lpor.201900449} {\bibfield  {journal} {\bibinfo  {journal} {Laser \& Photonics Reviews}\ }\textbf {\bibinfo {volume} {14}},\ \bibinfo {pages} {1900449} (\bibinfo {year} {2020})}\BibitemShut {NoStop}%
\bibitem [{\citenamefont {Spencer}\ \emph {et~al.}(2018)\citenamefont {Spencer}, \citenamefont {Drake}, \citenamefont {Briles}, \citenamefont {Stone}, \citenamefont {Sinclair}, \citenamefont {Fredrick}, \citenamefont {Li}, \citenamefont {Westly}, \citenamefont {Ilic}, \citenamefont {Bluestone}, \citenamefont {Volet}, \citenamefont {Komljenovic}, \citenamefont {Chang}, \citenamefont {Lee}, \citenamefont {Oh}, \citenamefont {Suh}, \citenamefont {Yang}, \citenamefont {Pfeiffer}, \citenamefont {Kippenberg}, \citenamefont {Norberg}, \citenamefont {Theogarajan}, \citenamefont {Vahala}, \citenamefont {Newbury}, \citenamefont {Srinivasan}, \citenamefont {Bowers}, \citenamefont {Diddams},\ and\ \citenamefont {Papp}}]{Spencer2018}%
  \BibitemOpen
  \bibfield  {author} {\bibinfo {author} {\bibfnamefont {D.~T.}\ \bibnamefont {Spencer}}, \bibinfo {author} {\bibfnamefont {T.}~\bibnamefont {Drake}}, \bibinfo {author} {\bibfnamefont {T.~C.}\ \bibnamefont {Briles}}, \bibinfo {author} {\bibfnamefont {J.}~\bibnamefont {Stone}}, \bibinfo {author} {\bibfnamefont {L.~C.}\ \bibnamefont {Sinclair}}, \bibinfo {author} {\bibfnamefont {C.}~\bibnamefont {Fredrick}}, \bibinfo {author} {\bibfnamefont {Q.}~\bibnamefont {Li}}, \bibinfo {author} {\bibfnamefont {D.}~\bibnamefont {Westly}}, \bibinfo {author} {\bibfnamefont {B.~R.}\ \bibnamefont {Ilic}}, \bibinfo {author} {\bibfnamefont {A.}~\bibnamefont {Bluestone}}, \bibinfo {author} {\bibfnamefont {N.}~\bibnamefont {Volet}}, \bibinfo {author} {\bibfnamefont {T.}~\bibnamefont {Komljenovic}}, \bibinfo {author} {\bibfnamefont {L.}~\bibnamefont {Chang}}, \bibinfo {author} {\bibfnamefont {S.~H.}\ \bibnamefont {Lee}}, \bibinfo {author} {\bibfnamefont {D.~Y.}\ \bibnamefont {Oh}}, \bibinfo {author} {\bibfnamefont {M.-G.}\
  \bibnamefont {Suh}}, \bibinfo {author} {\bibfnamefont {K.~Y.}\ \bibnamefont {Yang}}, \bibinfo {author} {\bibfnamefont {M.~H.~P.}\ \bibnamefont {Pfeiffer}}, \bibinfo {author} {\bibfnamefont {T.~J.}\ \bibnamefont {Kippenberg}}, \bibinfo {author} {\bibfnamefont {E.}~\bibnamefont {Norberg}}, \bibinfo {author} {\bibfnamefont {L.}~\bibnamefont {Theogarajan}}, \bibinfo {author} {\bibfnamefont {K.}~\bibnamefont {Vahala}}, \bibinfo {author} {\bibfnamefont {N.~R.}\ \bibnamefont {Newbury}}, \bibinfo {author} {\bibfnamefont {K.}~\bibnamefont {Srinivasan}}, \bibinfo {author} {\bibfnamefont {J.~E.}\ \bibnamefont {Bowers}}, \bibinfo {author} {\bibfnamefont {S.~A.}\ \bibnamefont {Diddams}},\ and\ \bibinfo {author} {\bibfnamefont {S.~B.}\ \bibnamefont {Papp}},\ }\href {https://doi.org/10.1038/s41586-018-0065-7} {\bibfield  {journal} {\bibinfo  {journal} {Nature}\ }\textbf {\bibinfo {volume} {557}},\ \bibinfo {pages} {81} (\bibinfo {year} {2018})}\BibitemShut {NoStop}%
\bibitem [{\citenamefont {Marin-Palomo}\ \emph {et~al.}(2017)\citenamefont {Marin-Palomo}, \citenamefont {Kemal}, \citenamefont {Karpov}, \citenamefont {Kordts}, \citenamefont {Pfeifle}, \citenamefont {Pfeiffer}, \citenamefont {Trocha}, \citenamefont {Wolf}, \citenamefont {Brasch}, \citenamefont {Anderson}, \citenamefont {Rosenberger}, \citenamefont {Vijayan}, \citenamefont {Freude}, \citenamefont {Kippenberg},\ and\ \citenamefont {Koos}}]{Marin-Palomo2017}%
  \BibitemOpen
  \bibfield  {author} {\bibinfo {author} {\bibfnamefont {P.}~\bibnamefont {Marin-Palomo}}, \bibinfo {author} {\bibfnamefont {J.~N.}\ \bibnamefont {Kemal}}, \bibinfo {author} {\bibfnamefont {M.}~\bibnamefont {Karpov}}, \bibinfo {author} {\bibfnamefont {A.}~\bibnamefont {Kordts}}, \bibinfo {author} {\bibfnamefont {J.}~\bibnamefont {Pfeifle}}, \bibinfo {author} {\bibfnamefont {M.~H.~P.}\ \bibnamefont {Pfeiffer}}, \bibinfo {author} {\bibfnamefont {P.}~\bibnamefont {Trocha}}, \bibinfo {author} {\bibfnamefont {S.}~\bibnamefont {Wolf}}, \bibinfo {author} {\bibfnamefont {V.}~\bibnamefont {Brasch}}, \bibinfo {author} {\bibfnamefont {M.~H.}\ \bibnamefont {Anderson}}, \bibinfo {author} {\bibfnamefont {R.}~\bibnamefont {Rosenberger}}, \bibinfo {author} {\bibfnamefont {K.}~\bibnamefont {Vijayan}}, \bibinfo {author} {\bibfnamefont {W.}~\bibnamefont {Freude}}, \bibinfo {author} {\bibfnamefont {T.~J.}\ \bibnamefont {Kippenberg}},\ and\ \bibinfo {author} {\bibfnamefont {C.}~\bibnamefont {Koos}},\ }\href
  {https://doi.org/10.1038/nature22387} {\bibfield  {journal} {\bibinfo  {journal} {Nature}\ }\textbf {\bibinfo {volume} {546}},\ \bibinfo {pages} {274} (\bibinfo {year} {2017})}\BibitemShut {NoStop}%
\bibitem [{\citenamefont {Rizzo}\ \emph {et~al.}(2023)\citenamefont {Rizzo}, \citenamefont {Novick}, \citenamefont {Gopal}, \citenamefont {Kim}, \citenamefont {Ji}, \citenamefont {Daudlin}, \citenamefont {Okawachi}, \citenamefont {Cheng}, \citenamefont {Lipson}, \citenamefont {Gaeta},\ and\ \citenamefont {Bergman}}]{Rizzo2023}%
  \BibitemOpen
  \bibfield  {author} {\bibinfo {author} {\bibfnamefont {A.}~\bibnamefont {Rizzo}}, \bibinfo {author} {\bibfnamefont {A.}~\bibnamefont {Novick}}, \bibinfo {author} {\bibfnamefont {V.}~\bibnamefont {Gopal}}, \bibinfo {author} {\bibfnamefont {B.~Y.}\ \bibnamefont {Kim}}, \bibinfo {author} {\bibfnamefont {X.}~\bibnamefont {Ji}}, \bibinfo {author} {\bibfnamefont {S.}~\bibnamefont {Daudlin}}, \bibinfo {author} {\bibfnamefont {Y.}~\bibnamefont {Okawachi}}, \bibinfo {author} {\bibfnamefont {Q.}~\bibnamefont {Cheng}}, \bibinfo {author} {\bibfnamefont {M.}~\bibnamefont {Lipson}}, \bibinfo {author} {\bibfnamefont {A.~L.}\ \bibnamefont {Gaeta}},\ and\ \bibinfo {author} {\bibfnamefont {K.}~\bibnamefont {Bergman}},\ }\href {https://doi.org/10.1038/s41566-023-01244-7} {\bibfield  {journal} {\bibinfo  {journal} {Nature Photonics}\ }\textbf {\bibinfo {volume} {17}},\ \bibinfo {pages} {781} (\bibinfo {year} {2023})}\BibitemShut {NoStop}%
\bibitem [{\citenamefont {Newman}\ \emph {et~al.}(2019)\citenamefont {Newman}, \citenamefont {Maurice}, \citenamefont {Drake}, \citenamefont {Stone}, \citenamefont {Briles}, \citenamefont {Spencer}, \citenamefont {Fredrick}, \citenamefont {Li}, \citenamefont {Westly}, \citenamefont {Ilic}, \citenamefont {Shen}, \citenamefont {Suh}, \citenamefont {Yang}, \citenamefont {Johnson}, \citenamefont {Johnson}, \citenamefont {Hollberg}, \citenamefont {Vahala}, \citenamefont {Srinivasan}, \citenamefont {Diddams}, \citenamefont {Kitching}, \citenamefont {Papp},\ and\ \citenamefont {Hummon}}]{newman_architecture_2019}%
  \BibitemOpen
  \bibfield  {author} {\bibinfo {author} {\bibfnamefont {Z.~L.}\ \bibnamefont {Newman}}, \bibinfo {author} {\bibfnamefont {V.}~\bibnamefont {Maurice}}, \bibinfo {author} {\bibfnamefont {T.}~\bibnamefont {Drake}}, \bibinfo {author} {\bibfnamefont {J.~R.}\ \bibnamefont {Stone}}, \bibinfo {author} {\bibfnamefont {T.~C.}\ \bibnamefont {Briles}}, \bibinfo {author} {\bibfnamefont {D.~T.}\ \bibnamefont {Spencer}}, \bibinfo {author} {\bibfnamefont {C.}~\bibnamefont {Fredrick}}, \bibinfo {author} {\bibfnamefont {Q.}~\bibnamefont {Li}}, \bibinfo {author} {\bibfnamefont {D.}~\bibnamefont {Westly}}, \bibinfo {author} {\bibfnamefont {B.~R.}\ \bibnamefont {Ilic}}, \bibinfo {author} {\bibfnamefont {B.}~\bibnamefont {Shen}}, \bibinfo {author} {\bibfnamefont {M.-G.}\ \bibnamefont {Suh}}, \bibinfo {author} {\bibfnamefont {K.~Y.}\ \bibnamefont {Yang}}, \bibinfo {author} {\bibfnamefont {C.}~\bibnamefont {Johnson}}, \bibinfo {author} {\bibfnamefont {D.~M.~S.}\ \bibnamefont {Johnson}}, \bibinfo {author} {\bibfnamefont
  {L.}~\bibnamefont {Hollberg}}, \bibinfo {author} {\bibfnamefont {K.~J.}\ \bibnamefont {Vahala}}, \bibinfo {author} {\bibfnamefont {K.}~\bibnamefont {Srinivasan}}, \bibinfo {author} {\bibfnamefont {S.~A.}\ \bibnamefont {Diddams}}, \bibinfo {author} {\bibfnamefont {J.}~\bibnamefont {Kitching}}, \bibinfo {author} {\bibfnamefont {S.~B.}\ \bibnamefont {Papp}},\ and\ \bibinfo {author} {\bibfnamefont {M.~T.}\ \bibnamefont {Hummon}},\ }\href {https://doi.org/10.1364/OPTICA.6.000680} {\bibfield  {journal} {\bibinfo  {journal} {Optica}\ }\textbf {\bibinfo {volume} {6}},\ \bibinfo {pages} {680} (\bibinfo {year} {2019})},\ \bibinfo {note} {publisher: Optica Publishing Group}\BibitemShut {NoStop}%
\bibitem [{\citenamefont {Pupeza}\ \emph {et~al.}(2021)\citenamefont {Pupeza}, \citenamefont {Zhang}, \citenamefont {H{\"o}gner},\ and\ \citenamefont {Ye}}]{Pupeza2021}%
  \BibitemOpen
  \bibfield  {author} {\bibinfo {author} {\bibfnamefont {I.}~\bibnamefont {Pupeza}}, \bibinfo {author} {\bibfnamefont {C.}~\bibnamefont {Zhang}}, \bibinfo {author} {\bibfnamefont {M.}~\bibnamefont {H{\"o}gner}},\ and\ \bibinfo {author} {\bibfnamefont {J.}~\bibnamefont {Ye}},\ }\href {https://doi.org/10.1038/s41566-020-00741-3} {\bibfield  {journal} {\bibinfo  {journal} {Nature Photonics}\ }\textbf {\bibinfo {volume} {15}},\ \bibinfo {pages} {175} (\bibinfo {year} {2021})}\BibitemShut {NoStop}%
\bibitem [{\citenamefont {Yu}\ \emph {et~al.}(2019{\natexlab{a}})\citenamefont {Yu}, \citenamefont {Okawachi}, \citenamefont {Griffith}, \citenamefont {Lipson},\ and\ \citenamefont {Gaeta}}]{Yu:19}%
  \BibitemOpen
  \bibfield  {author} {\bibinfo {author} {\bibfnamefont {M.}~\bibnamefont {Yu}}, \bibinfo {author} {\bibfnamefont {Y.}~\bibnamefont {Okawachi}}, \bibinfo {author} {\bibfnamefont {A.~G.}\ \bibnamefont {Griffith}}, \bibinfo {author} {\bibfnamefont {M.}~\bibnamefont {Lipson}},\ and\ \bibinfo {author} {\bibfnamefont {A.~L.}\ \bibnamefont {Gaeta}},\ }\href {https://doi.org/10.1364/OL.44.004259} {\bibfield  {journal} {\bibinfo  {journal} {Opt. Lett.}\ }\textbf {\bibinfo {volume} {44}},\ \bibinfo {pages} {4259} (\bibinfo {year} {2019}{\natexlab{a}})}\BibitemShut {NoStop}%
\bibitem [{\citenamefont {Marchand}\ \emph {et~al.}(2021)\citenamefont {Marchand}, \citenamefont {Riemensberger}, \citenamefont {Skehan}, \citenamefont {Ho}, \citenamefont {Pfeiffer}, \citenamefont {Liu}, \citenamefont {Hauger}, \citenamefont {Lasser},\ and\ \citenamefont {Kippenberg}}]{Marchand2021}%
  \BibitemOpen
  \bibfield  {author} {\bibinfo {author} {\bibfnamefont {P.~J.}\ \bibnamefont {Marchand}}, \bibinfo {author} {\bibfnamefont {J.}~\bibnamefont {Riemensberger}}, \bibinfo {author} {\bibfnamefont {J.~C.}\ \bibnamefont {Skehan}}, \bibinfo {author} {\bibfnamefont {J.-J.}\ \bibnamefont {Ho}}, \bibinfo {author} {\bibfnamefont {M.~H.~P.}\ \bibnamefont {Pfeiffer}}, \bibinfo {author} {\bibfnamefont {J.}~\bibnamefont {Liu}}, \bibinfo {author} {\bibfnamefont {C.}~\bibnamefont {Hauger}}, \bibinfo {author} {\bibfnamefont {T.}~\bibnamefont {Lasser}},\ and\ \bibinfo {author} {\bibfnamefont {T.~J.}\ \bibnamefont {Kippenberg}},\ }\href {https://doi.org/10.1038/s41467-020-20404-9} {\bibfield  {journal} {\bibinfo  {journal} {Nature Communications}\ }\textbf {\bibinfo {volume} {12}},\ \bibinfo {pages} {427} (\bibinfo {year} {2021})}\BibitemShut {NoStop}%
\bibitem [{\citenamefont {Dudley}(2006)}]{dudley_supercontinuum_2006}%
  \BibitemOpen
  \bibfield  {author} {\bibinfo {author} {\bibfnamefont {J.~M.}\ \bibnamefont {Dudley}},\ }\href {https://doi.org/10.1103/RevModPhys.78.1135} {\bibfield  {journal} {\bibinfo  {journal} {Reviews of Modern Physics}\ }\textbf {\bibinfo {volume} {78}},\ \bibinfo {pages} {1135} (\bibinfo {year} {2006})}\BibitemShut {NoStop}%
\bibitem [{\citenamefont {Travers}\ \emph {et~al.}(2010)\citenamefont {Travers}, \citenamefont {Frosz},\ and\ \citenamefont {Dudley}}]{Dudley_2010}%
  \BibitemOpen
  \bibfield  {author} {\bibinfo {author} {\bibfnamefont {J.~C.}\ \bibnamefont {Travers}}, \bibinfo {author} {\bibfnamefont {M.~H.}\ \bibnamefont {Frosz}},\ and\ \bibinfo {author} {\bibfnamefont {J.~M.}\ \bibnamefont {Dudley}},\ }\bibinfo {title} {Nonlinear fibre optics overview},\ in\ \href@noop {} {\emph {\bibinfo {booktitle} {Supercontinuum Generation in Optical Fibers}}},\ \bibinfo {editor} {edited by\ \bibinfo {editor} {\bibfnamefont {J.~M.}\ \bibnamefont {Dudley}}\ and\ \bibinfo {editor} {\bibfnamefont {J.~R.}\ \bibnamefont {Taylor}}}\ (\bibinfo  {publisher} {Cambridge University Press},\ \bibinfo {year} {2010})\ pp.\ \bibinfo {pages} {32--51}\BibitemShut {NoStop}%
\bibitem [{\citenamefont {Yue}\ \emph {et~al.}(2024)\citenamefont {Yue}, \citenamefont {Fang}, \citenamefont {Geng},\ and\ \citenamefont {Bao}}]{Yue2024IntegratedSCG}%
  \BibitemOpen
  \bibfield  {author} {\bibinfo {author} {\bibfnamefont {Y.}~\bibnamefont {Yue}}, \bibinfo {author} {\bibfnamefont {Y.}~\bibnamefont {Fang}}, \bibinfo {author} {\bibfnamefont {W.}~\bibnamefont {Geng}},\ and\ \bibinfo {author} {\bibfnamefont {C.}~\bibnamefont {Bao}},\ }\href {https://doi.org/10.1007/978-981-97-6584-3} {\emph {\bibinfo {title} {Integrated Optical Supercontinuum Generation: Physics, Advances, and Applications}}},\ \bibinfo {edition} {1st}\ ed.,\ Advances in Optics and Optoelectronics\ (\bibinfo  {publisher} {Springer Singapore},\ \bibinfo {year} {2024})\ pp.\ \bibinfo {pages} {VII, 198}\BibitemShut {NoStop}%
\bibitem [{\citenamefont {Cao}\ \emph {et~al.}(2022)\citenamefont {Cao}, \citenamefont {Sohn}, \citenamefont {Gao}, \citenamefont {Xing}, \citenamefont {Chen}, \citenamefont {Ng},\ and\ \citenamefont {Tan}}]{Cao2022}%
  \BibitemOpen
  \bibfield  {author} {\bibinfo {author} {\bibfnamefont {Y.}~\bibnamefont {Cao}}, \bibinfo {author} {\bibfnamefont {B.-U.}\ \bibnamefont {Sohn}}, \bibinfo {author} {\bibfnamefont {H.}~\bibnamefont {Gao}}, \bibinfo {author} {\bibfnamefont {P.}~\bibnamefont {Xing}}, \bibinfo {author} {\bibfnamefont {G.~F.~R.}\ \bibnamefont {Chen}}, \bibinfo {author} {\bibfnamefont {D.~K.~T.}\ \bibnamefont {Ng}},\ and\ \bibinfo {author} {\bibfnamefont {D.~T.~H.}\ \bibnamefont {Tan}},\ }\href {https://doi.org/10.1038/s41598-022-13734-9} {\bibfield  {journal} {\bibinfo  {journal} {Scientific Reports}\ }\textbf {\bibinfo {volume} {12}},\ \bibinfo {pages} {9487} (\bibinfo {year} {2022})}\BibitemShut {NoStop}%
\bibitem [{\citenamefont {Lu}\ \emph {et~al.}(2019)\citenamefont {Lu}, \citenamefont {Surya}, \citenamefont {Liu}, \citenamefont {Xu},\ and\ \citenamefont {Tang}}]{Lu:19_chi3}%
  \BibitemOpen
  \bibfield  {author} {\bibinfo {author} {\bibfnamefont {J.}~\bibnamefont {Lu}}, \bibinfo {author} {\bibfnamefont {J.~B.}\ \bibnamefont {Surya}}, \bibinfo {author} {\bibfnamefont {X.}~\bibnamefont {Liu}}, \bibinfo {author} {\bibfnamefont {Y.}~\bibnamefont {Xu}},\ and\ \bibinfo {author} {\bibfnamefont {H.~X.}\ \bibnamefont {Tang}},\ }\href {https://doi.org/10.1364/OL.44.001492} {\bibfield  {journal} {\bibinfo  {journal} {Opt. Lett.}\ }\textbf {\bibinfo {volume} {44}},\ \bibinfo {pages} {1492} (\bibinfo {year} {2019})}\BibitemShut {NoStop}%
\bibitem [{\citenamefont {Yu}\ \emph {et~al.}(2019{\natexlab{b}})\citenamefont {Yu}, \citenamefont {Desiatov}, \citenamefont {Okawachi}, \citenamefont {Gaeta},\ and\ \citenamefont {Lon\v{c}ar}}]{Yu:19_chi3}%
  \BibitemOpen
  \bibfield  {author} {\bibinfo {author} {\bibfnamefont {M.}~\bibnamefont {Yu}}, \bibinfo {author} {\bibfnamefont {B.}~\bibnamefont {Desiatov}}, \bibinfo {author} {\bibfnamefont {Y.}~\bibnamefont {Okawachi}}, \bibinfo {author} {\bibfnamefont {A.~L.}\ \bibnamefont {Gaeta}},\ and\ \bibinfo {author} {\bibfnamefont {M.}~\bibnamefont {Lon\v{c}ar}},\ }\href {https://doi.org/10.1364/OL.44.001222} {\bibfield  {journal} {\bibinfo  {journal} {Opt. Lett.}\ }\textbf {\bibinfo {volume} {44}},\ \bibinfo {pages} {1222} (\bibinfo {year} {2019}{\natexlab{b}})}\BibitemShut {NoStop}%
\bibitem [{\citenamefont {Gao}\ \emph {et~al.}(2025)\citenamefont {Gao}, \citenamefont {Sun}, \citenamefont {Rebolledo-Salgado}, \citenamefont {Laer}, \citenamefont {Torres-Company},\ and\ \citenamefont {Schroder}}]{gao2025_chi3}%
  \BibitemOpen
  \bibfield  {author} {\bibinfo {author} {\bibfnamefont {Y.}~\bibnamefont {Gao}}, \bibinfo {author} {\bibfnamefont {Y.}~\bibnamefont {Sun}}, \bibinfo {author} {\bibfnamefont {I.}~\bibnamefont {Rebolledo-Salgado}}, \bibinfo {author} {\bibfnamefont {R.~V.}\ \bibnamefont {Laer}}, \bibinfo {author} {\bibfnamefont {V.}~\bibnamefont {Torres-Company}},\ and\ \bibinfo {author} {\bibfnamefont {J.}~\bibnamefont {Schroder}},\ }\href {https://arxiv.org/abs/2501.18341} {\bibinfo {title} {Tightly-confined and long z-cut lithium niobate waveguide with ultralow-loss}} (\bibinfo {year} {2025}),\ \Eprint {https://arxiv.org/abs/2501.18341} {arXiv:2501.18341 [physics.optics]} \BibitemShut {NoStop}%
\bibitem [{\citenamefont {Guo}\ \emph {et~al.}(2018)\citenamefont {Guo}, \citenamefont {Herkommer}, \citenamefont {Billat}, \citenamefont {Grassani}, \citenamefont {Zhang}, \citenamefont {Pfeiffer}, \citenamefont {Weng}, \citenamefont {Br{\`e}s},\ and\ \citenamefont {Kippenberg}}]{Guo2018}%
  \BibitemOpen
  \bibfield  {author} {\bibinfo {author} {\bibfnamefont {H.}~\bibnamefont {Guo}}, \bibinfo {author} {\bibfnamefont {C.}~\bibnamefont {Herkommer}}, \bibinfo {author} {\bibfnamefont {A.}~\bibnamefont {Billat}}, \bibinfo {author} {\bibfnamefont {D.}~\bibnamefont {Grassani}}, \bibinfo {author} {\bibfnamefont {C.}~\bibnamefont {Zhang}}, \bibinfo {author} {\bibfnamefont {M.~H.~P.}\ \bibnamefont {Pfeiffer}}, \bibinfo {author} {\bibfnamefont {W.}~\bibnamefont {Weng}}, \bibinfo {author} {\bibfnamefont {C.-S.}\ \bibnamefont {Br{\`e}s}},\ and\ \bibinfo {author} {\bibfnamefont {T.~J.}\ \bibnamefont {Kippenberg}},\ }\href {https://doi.org/10.1038/s41566-018-0144-1} {\bibfield  {journal} {\bibinfo  {journal} {Nature Photonics}\ }\textbf {\bibinfo {volume} {12}},\ \bibinfo {pages} {330} (\bibinfo {year} {2018})}\BibitemShut {NoStop}%
\bibitem [{\citenamefont {Yamada}\ \emph {et~al.}(1993)\citenamefont {Yamada}, \citenamefont {Nada}, \citenamefont {Saitoh},\ and\ \citenamefont {Watanabe}}]{yamada_1993_qpm}%
  \BibitemOpen
  \bibfield  {author} {\bibinfo {author} {\bibfnamefont {M.}~\bibnamefont {Yamada}}, \bibinfo {author} {\bibfnamefont {N.}~\bibnamefont {Nada}}, \bibinfo {author} {\bibfnamefont {M.}~\bibnamefont {Saitoh}},\ and\ \bibinfo {author} {\bibfnamefont {K.}~\bibnamefont {Watanabe}},\ }\href {https://doi.org/10.1063/1.108925} {\bibfield  {journal} {\bibinfo  {journal} {Applied Physics Letters}\ }\textbf {\bibinfo {volume} {62}},\ \bibinfo {pages} {435} (\bibinfo {year} {1993})}\BibitemShut {NoStop}%
\bibitem [{\citenamefont {Jankowski}\ \emph {et~al.}(2023)\citenamefont {Jankowski}, \citenamefont {Langrock}, \citenamefont {Desiatov}, \citenamefont {Lončar},\ and\ \citenamefont {Fejer}}]{Marc:23}%
  \BibitemOpen
  \bibfield  {author} {\bibinfo {author} {\bibfnamefont {M.}~\bibnamefont {Jankowski}}, \bibinfo {author} {\bibfnamefont {C.}~\bibnamefont {Langrock}}, \bibinfo {author} {\bibfnamefont {B.}~\bibnamefont {Desiatov}}, \bibinfo {author} {\bibfnamefont {M.}~\bibnamefont {Lončar}},\ and\ \bibinfo {author} {\bibfnamefont {M.~M.}\ \bibnamefont {Fejer}},\ }\href {https://doi.org/10.1063/5.0158926} {\bibfield  {journal} {\bibinfo  {journal} {APL Photonics}\ }\textbf {\bibinfo {volume} {8}},\ \bibinfo {pages} {116104} (\bibinfo {year} {2023})}\BibitemShut {NoStop}%
\bibitem [{\citenamefont {Phillips}\ \emph {et~al.}(2011)\citenamefont {Phillips}, \citenamefont {Langrock}, \citenamefont {Pelc}, \citenamefont {Fejer}, \citenamefont {Jiang}, \citenamefont {Fermann},\ and\ \citenamefont {Hartl}}]{Phillips2011_chi2}%
  \BibitemOpen
  \bibfield  {author} {\bibinfo {author} {\bibfnamefont {C.~R.}\ \bibnamefont {Phillips}}, \bibinfo {author} {\bibfnamefont {C.}~\bibnamefont {Langrock}}, \bibinfo {author} {\bibfnamefont {J.~S.}\ \bibnamefont {Pelc}}, \bibinfo {author} {\bibfnamefont {M.~M.}\ \bibnamefont {Fejer}}, \bibinfo {author} {\bibfnamefont {J.}~\bibnamefont {Jiang}}, \bibinfo {author} {\bibfnamefont {M.~E.}\ \bibnamefont {Fermann}},\ and\ \bibinfo {author} {\bibfnamefont {I.}~\bibnamefont {Hartl}},\ }\href@noop {} {\bibfield  {journal} {\bibinfo  {journal} {Opt. Lett.}\ }\textbf {\bibinfo {volume} {36}},\ \bibinfo {pages} {3912} (\bibinfo {year} {2011})}\BibitemShut {NoStop}%
\bibitem [{\citenamefont {Jankowski}\ \emph {et~al.}(2022)\citenamefont {Jankowski}, \citenamefont {Jornod}, \citenamefont {Langrock}, \citenamefont {Desiatov}, \citenamefont {Marandi}, \citenamefont {Lon\v{c}ar},\ and\ \citenamefont {Fejer}}]{Jankowski:22:quasi}%
  \BibitemOpen
  \bibfield  {author} {\bibinfo {author} {\bibfnamefont {M.}~\bibnamefont {Jankowski}}, \bibinfo {author} {\bibfnamefont {N.}~\bibnamefont {Jornod}}, \bibinfo {author} {\bibfnamefont {C.}~\bibnamefont {Langrock}}, \bibinfo {author} {\bibfnamefont {B.}~\bibnamefont {Desiatov}}, \bibinfo {author} {\bibfnamefont {A.}~\bibnamefont {Marandi}}, \bibinfo {author} {\bibfnamefont {M.}~\bibnamefont {Lon\v{c}ar}},\ and\ \bibinfo {author} {\bibfnamefont {M.~M.}\ \bibnamefont {Fejer}},\ }\href {https://doi.org/10.1364/OPTICA.442550} {\bibfield  {journal} {\bibinfo  {journal} {Optica}\ }\textbf {\bibinfo {volume} {9}},\ \bibinfo {pages} {273} (\bibinfo {year} {2022})}\BibitemShut {NoStop}%
\bibitem [{\citenamefont {Kowligy}\ \emph {et~al.}(2018)\citenamefont {Kowligy}, \citenamefont {Lind}, \citenamefont {Hickstein}, \citenamefont {Carlson}, \citenamefont {Timmers}, \citenamefont {Nader}, \citenamefont {Cruz}, \citenamefont {Ycas}, \citenamefont {Papp},\ and\ \citenamefont {Diddams}}]{Kowligy:18}%
  \BibitemOpen
  \bibfield  {author} {\bibinfo {author} {\bibfnamefont {A.~S.}\ \bibnamefont {Kowligy}}, \bibinfo {author} {\bibfnamefont {A.}~\bibnamefont {Lind}}, \bibinfo {author} {\bibfnamefont {D.~D.}\ \bibnamefont {Hickstein}}, \bibinfo {author} {\bibfnamefont {D.~R.}\ \bibnamefont {Carlson}}, \bibinfo {author} {\bibfnamefont {H.}~\bibnamefont {Timmers}}, \bibinfo {author} {\bibfnamefont {N.}~\bibnamefont {Nader}}, \bibinfo {author} {\bibfnamefont {F.~C.}\ \bibnamefont {Cruz}}, \bibinfo {author} {\bibfnamefont {G.}~\bibnamefont {Ycas}}, \bibinfo {author} {\bibfnamefont {S.~B.}\ \bibnamefont {Papp}},\ and\ \bibinfo {author} {\bibfnamefont {S.~A.}\ \bibnamefont {Diddams}},\ }\href {https://doi.org/10.1364/OL.43.001678} {\bibfield  {journal} {\bibinfo  {journal} {Opt. Lett.}\ }\textbf {\bibinfo {volume} {43}},\ \bibinfo {pages} {1678} (\bibinfo {year} {2018})}\BibitemShut {NoStop}%
\bibitem [{\citenamefont {Wu}\ \emph {et~al.}(2024)\citenamefont {Wu}, \citenamefont {Ledezma}, \citenamefont {Fredrick}, \citenamefont {Sekhar}, \citenamefont {Sekine}, \citenamefont {Guo}, \citenamefont {Briggs}, \citenamefont {Marandi},\ and\ \citenamefont {Diddams}}]{Wu2024}%
  \BibitemOpen
  \bibfield  {author} {\bibinfo {author} {\bibfnamefont {T.-H.}\ \bibnamefont {Wu}}, \bibinfo {author} {\bibfnamefont {L.}~\bibnamefont {Ledezma}}, \bibinfo {author} {\bibfnamefont {C.}~\bibnamefont {Fredrick}}, \bibinfo {author} {\bibfnamefont {P.}~\bibnamefont {Sekhar}}, \bibinfo {author} {\bibfnamefont {R.}~\bibnamefont {Sekine}}, \bibinfo {author} {\bibfnamefont {Q.}~\bibnamefont {Guo}}, \bibinfo {author} {\bibfnamefont {R.~M.}\ \bibnamefont {Briggs}}, \bibinfo {author} {\bibfnamefont {A.}~\bibnamefont {Marandi}},\ and\ \bibinfo {author} {\bibfnamefont {S.~A.}\ \bibnamefont {Diddams}},\ }\href {https://doi.org/10.1038/s41566-023-01364-0} {\bibfield  {journal} {\bibinfo  {journal} {Nature Photonics}\ }\textbf {\bibinfo {volume} {18}},\ \bibinfo {pages} {218} (\bibinfo {year} {2024})}\BibitemShut {NoStop}%
\bibitem [{\citenamefont {Hamrouni}\ \emph {et~al.}(2024)\citenamefont {Hamrouni}, \citenamefont {Jankowski}, \citenamefont {Hwang}, \citenamefont {Flemens}, \citenamefont {Mishra}, \citenamefont {Langrock}, \citenamefont {Safavi-Naeini}, \citenamefont {Fejer},\ and\ \citenamefont {S\"{u}dmeyer}}]{Hamrouni:24}%
  \BibitemOpen
  \bibfield  {author} {\bibinfo {author} {\bibfnamefont {M.}~\bibnamefont {Hamrouni}}, \bibinfo {author} {\bibfnamefont {M.}~\bibnamefont {Jankowski}}, \bibinfo {author} {\bibfnamefont {A.~Y.}\ \bibnamefont {Hwang}}, \bibinfo {author} {\bibfnamefont {N.}~\bibnamefont {Flemens}}, \bibinfo {author} {\bibfnamefont {J.}~\bibnamefont {Mishra}}, \bibinfo {author} {\bibfnamefont {C.}~\bibnamefont {Langrock}}, \bibinfo {author} {\bibfnamefont {A.~H.}\ \bibnamefont {Safavi-Naeini}}, \bibinfo {author} {\bibfnamefont {M.~M.}\ \bibnamefont {Fejer}},\ and\ \bibinfo {author} {\bibfnamefont {T.}~\bibnamefont {S\"{u}dmeyer}},\ }\href {https://doi.org/10.1364/OE.514649} {\bibfield  {journal} {\bibinfo  {journal} {Optics Express}\ }\textbf {\bibinfo {volume} {32}},\ \bibinfo {pages} {12004} (\bibinfo {year} {2024})}\BibitemShut {NoStop}%
\bibitem [{\citenamefont {Jankowski}\ \emph {et~al.}(2021)\citenamefont {Jankowski}, \citenamefont {Mishra},\ and\ \citenamefont {Fejer}}]{Jankowski2021}%
  \BibitemOpen
  \bibfield  {author} {\bibinfo {author} {\bibfnamefont {M.}~\bibnamefont {Jankowski}}, \bibinfo {author} {\bibfnamefont {J.}~\bibnamefont {Mishra}},\ and\ \bibinfo {author} {\bibfnamefont {M.~M.}\ \bibnamefont {Fejer}},\ }\href {https://doi.org/10.1088/2515-7647/ac1729} {\bibfield  {journal} {\bibinfo  {journal} {Journal of Physics: Photonics}\ }\textbf {\bibinfo {volume} {3}},\ \bibinfo {pages} {042005} (\bibinfo {year} {2021})}\BibitemShut {NoStop}%
\bibitem [{\citenamefont {Ledezma}\ \emph {et~al.}(2022)\citenamefont {Ledezma}, \citenamefont {Sekine}, \citenamefont {Guo}, \citenamefont {Nehra}, \citenamefont {Jahani},\ and\ \citenamefont {Marandi}}]{Ledezma:22}%
  \BibitemOpen
  \bibfield  {author} {\bibinfo {author} {\bibfnamefont {L.}~\bibnamefont {Ledezma}}, \bibinfo {author} {\bibfnamefont {R.}~\bibnamefont {Sekine}}, \bibinfo {author} {\bibfnamefont {Q.}~\bibnamefont {Guo}}, \bibinfo {author} {\bibfnamefont {R.}~\bibnamefont {Nehra}}, \bibinfo {author} {\bibfnamefont {S.}~\bibnamefont {Jahani}},\ and\ \bibinfo {author} {\bibfnamefont {A.}~\bibnamefont {Marandi}},\ }\href {https://doi.org/10.1364/OPTICA.442332} {\bibfield  {journal} {\bibinfo  {journal} {Optica}\ }\textbf {\bibinfo {volume} {9}},\ \bibinfo {pages} {303} (\bibinfo {year} {2022})}\BibitemShut {NoStop}%
\bibitem [{\citenamefont {Fan}\ \emph {et~al.}(2025)\citenamefont {Fan}, \citenamefont {Ayhan}, \citenamefont {Wildi}, \citenamefont {Volkov}, \citenamefont {Seer}, \citenamefont {Ludwig}, \citenamefont {Voumard}, \citenamefont {Brodschelm}, \citenamefont {Brasch}, \citenamefont {Villanueva},\ and\ \citenamefont {Herr}}]{fan2025_chirpchi2}%
  \BibitemOpen
  \bibfield  {author} {\bibinfo {author} {\bibfnamefont {W.}~\bibnamefont {Fan}}, \bibinfo {author} {\bibfnamefont {F.}~\bibnamefont {Ayhan}}, \bibinfo {author} {\bibfnamefont {T.}~\bibnamefont {Wildi}}, \bibinfo {author} {\bibfnamefont {M.}~\bibnamefont {Volkov}}, \bibinfo {author} {\bibfnamefont {A.}~\bibnamefont {Seer}}, \bibinfo {author} {\bibfnamefont {M.}~\bibnamefont {Ludwig}}, \bibinfo {author} {\bibfnamefont {T.}~\bibnamefont {Voumard}}, \bibinfo {author} {\bibfnamefont {A.}~\bibnamefont {Brodschelm}}, \bibinfo {author} {\bibfnamefont {V.}~\bibnamefont {Brasch}}, \bibinfo {author} {\bibfnamefont {G.~L.}\ \bibnamefont {Villanueva}},\ and\ \bibinfo {author} {\bibfnamefont {T.}~\bibnamefont {Herr}},\ }\href {https://arxiv.org/abs/2503.04468} {\bibinfo {title} {Spectral dynamics in broadband frequency combs with overlapping harmonics}} (\bibinfo {year} {2025}),\ \Eprint {https://arxiv.org/abs/2503.04468} {arXiv:2503.04468 [physics.optics]} \BibitemShut {NoStop}%
\bibitem [{\citenamefont {Mishra}\ \emph {et~al.}(2021)\citenamefont {Mishra}, \citenamefont {McKenna}, \citenamefont {Ng}, \citenamefont {Stokowski}, \citenamefont {Jankowski}, \citenamefont {Langrock}, \citenamefont {Heydari}, \citenamefont {Mabuchi}, \citenamefont {Fejer},\ and\ \citenamefont {Safavi-Naeini}}]{Mishra:21}%
  \BibitemOpen
  \bibfield  {author} {\bibinfo {author} {\bibfnamefont {J.}~\bibnamefont {Mishra}}, \bibinfo {author} {\bibfnamefont {T.~P.}\ \bibnamefont {McKenna}}, \bibinfo {author} {\bibfnamefont {E.}~\bibnamefont {Ng}}, \bibinfo {author} {\bibfnamefont {H.~S.}\ \bibnamefont {Stokowski}}, \bibinfo {author} {\bibfnamefont {M.}~\bibnamefont {Jankowski}}, \bibinfo {author} {\bibfnamefont {C.}~\bibnamefont {Langrock}}, \bibinfo {author} {\bibfnamefont {D.}~\bibnamefont {Heydari}}, \bibinfo {author} {\bibfnamefont {H.}~\bibnamefont {Mabuchi}}, \bibinfo {author} {\bibfnamefont {M.~M.}\ \bibnamefont {Fejer}},\ and\ \bibinfo {author} {\bibfnamefont {A.~H.}\ \bibnamefont {Safavi-Naeini}},\ }\href {https://doi.org/10.1364/OPTICA.427428} {\bibfield  {journal} {\bibinfo  {journal} {Optica}\ }\textbf {\bibinfo {volume} {8}},\ \bibinfo {pages} {921} (\bibinfo {year} {2021})}\BibitemShut {NoStop}%
\bibitem [{\citenamefont {Mishra}\ \emph {et~al.}(2022)\citenamefont {Mishra}, \citenamefont {Jankowski}, \citenamefont {Hwang}, \citenamefont {Stokowski}, \citenamefont {McKenna}, \citenamefont {Langrock}, \citenamefont {Ng}, \citenamefont {Heydari}, \citenamefont {Mabuchi}, \citenamefont {Safavi-Naeini},\ and\ \citenamefont {Fejer}}]{Mishra:22}%
  \BibitemOpen
  \bibfield  {author} {\bibinfo {author} {\bibfnamefont {J.}~\bibnamefont {Mishra}}, \bibinfo {author} {\bibfnamefont {M.}~\bibnamefont {Jankowski}}, \bibinfo {author} {\bibfnamefont {A.~Y.}\ \bibnamefont {Hwang}}, \bibinfo {author} {\bibfnamefont {H.~S.}\ \bibnamefont {Stokowski}}, \bibinfo {author} {\bibfnamefont {T.~P.}\ \bibnamefont {McKenna}}, \bibinfo {author} {\bibfnamefont {C.}~\bibnamefont {Langrock}}, \bibinfo {author} {\bibfnamefont {E.}~\bibnamefont {Ng}}, \bibinfo {author} {\bibfnamefont {D.}~\bibnamefont {Heydari}}, \bibinfo {author} {\bibfnamefont {H.}~\bibnamefont {Mabuchi}}, \bibinfo {author} {\bibfnamefont {A.~H.}\ \bibnamefont {Safavi-Naeini}},\ and\ \bibinfo {author} {\bibfnamefont {M.~M.}\ \bibnamefont {Fejer}},\ }\href {https://doi.org/10.1364/OE.467580} {\bibfield  {journal} {\bibinfo  {journal} {Opt. Express}\ }\textbf {\bibinfo {volume} {30}},\ \bibinfo {pages} {32752} (\bibinfo {year} {2022})}\BibitemShut {NoStop}%
\bibitem [{\citenamefont {Xiong}\ \emph {et~al.}(2024)\citenamefont {Xiong}, \citenamefont {Cao}, \citenamefont {Li}, \citenamefont {Yao}, \citenamefont {Zhang}, \citenamefont {Yao}, \citenamefont {Bao}, \citenamefont {Huang}, \citenamefont {Yu},\ and\ \citenamefont {Dai}}]{Xiong24_cascadePPLN}%
  \BibitemOpen
  \bibfield  {author} {\bibinfo {author} {\bibfnamefont {H.}~\bibnamefont {Xiong}}, \bibinfo {author} {\bibfnamefont {H.}~\bibnamefont {Cao}}, \bibinfo {author} {\bibfnamefont {M.}~\bibnamefont {Li}}, \bibinfo {author} {\bibfnamefont {Q.}~\bibnamefont {Yao}}, \bibinfo {author} {\bibfnamefont {M.}~\bibnamefont {Zhang}}, \bibinfo {author} {\bibfnamefont {X.}~\bibnamefont {Yao}}, \bibinfo {author} {\bibfnamefont {Y.}~\bibnamefont {Bao}}, \bibinfo {author} {\bibfnamefont {F.}~\bibnamefont {Huang}}, \bibinfo {author} {\bibfnamefont {Z.}~\bibnamefont {Yu}},\ and\ \bibinfo {author} {\bibfnamefont {D.}~\bibnamefont {Dai}},\ }\href {https://doi.org/10.1109/JLT.2024.3380364} {\bibfield  {journal} {\bibinfo  {journal} {Journal of Lightwave Technology}\ }\textbf {\bibinfo {volume} {42}},\ \bibinfo {pages} {4892} (\bibinfo {year} {2024})}\BibitemShut {NoStop}%
\bibitem [{\citenamefont {Shen}\ \emph {et~al.}(2025)\citenamefont {Shen}, \citenamefont {Yang}, \citenamefont {Li}, \citenamefont {Wang}, \citenamefont {Zhang}, \citenamefont {Zhang}, \citenamefont {Pan}, \citenamefont {Chang},\ and\ \citenamefont {Xin}}]{Shen:25}%
  \BibitemOpen
  \bibfield  {author} {\bibinfo {author} {\bibfnamefont {Y.}~\bibnamefont {Shen}}, \bibinfo {author} {\bibfnamefont {J.}~\bibnamefont {Yang}}, \bibinfo {author} {\bibfnamefont {M.}~\bibnamefont {Li}}, \bibinfo {author} {\bibfnamefont {T.}~\bibnamefont {Wang}}, \bibinfo {author} {\bibfnamefont {Y.}~\bibnamefont {Zhang}}, \bibinfo {author} {\bibfnamefont {K.}~\bibnamefont {Zhang}}, \bibinfo {author} {\bibfnamefont {D.}~\bibnamefont {Pan}}, \bibinfo {author} {\bibfnamefont {G.}~\bibnamefont {Chang}},\ and\ \bibinfo {author} {\bibfnamefont {M.}~\bibnamefont {Xin}},\ }\href {https://doi.org/10.1364/OE.559726} {\bibfield  {journal} {\bibinfo  {journal} {Opt. Express}\ }\textbf {\bibinfo {volume} {33}},\ \bibinfo {pages} {20129} (\bibinfo {year} {2025})}\BibitemShut {NoStop}%
\bibitem [{\citenamefont {Conforti}\ \emph {et~al.}(2010)\citenamefont {Conforti}, \citenamefont {Baronio},\ and\ \citenamefont {De~Angelis}}]{Conforti2011}%
  \BibitemOpen
  \bibfield  {author} {\bibinfo {author} {\bibfnamefont {M.}~\bibnamefont {Conforti}}, \bibinfo {author} {\bibfnamefont {F.}~\bibnamefont {Baronio}},\ and\ \bibinfo {author} {\bibfnamefont {C.}~\bibnamefont {De~Angelis}},\ }\href {https://doi.org/10.1103/PhysRevA.81.053841} {\bibfield  {journal} {\bibinfo  {journal} {Phys. Rev. A}\ }\textbf {\bibinfo {volume} {81}},\ \bibinfo {pages} {053841} (\bibinfo {year} {2010})}\BibitemShut {NoStop}%
\bibitem [{\citenamefont {{Gordon}}\ \emph {et~al.}(2022)\citenamefont {{Gordon}}, \citenamefont {{Rothman}}, \citenamefont {{Hargreaves}}, \citenamefont {{Hashemi}}, \citenamefont {{Karlovets}}, \citenamefont {{Skinner}}, \citenamefont {{Conway}}, \citenamefont {{Hill}}, \citenamefont {{Kochanov}}, \citenamefont {{Tan}}, \citenamefont {{Wcis{\l}o}}, \citenamefont {{Finenko}}, \citenamefont {{Nelson}}, \citenamefont {{Bernath}}, \citenamefont {{Birk}}, \citenamefont {{Boudon}}, \citenamefont {{Campargue}}, \citenamefont {{Chance}}, \citenamefont {{Coustenis}}, \citenamefont {{Drouin}}, \citenamefont {{Flaud}}, \citenamefont {{Gamache}}, \citenamefont {{Hodges}}, \citenamefont {{Jacquemart}}, \citenamefont {{Mlawer}}, \citenamefont {{Nikitin}}, \citenamefont {{Perevalov}}, \citenamefont {{Rotger}}, \citenamefont {{Tennyson}}, \citenamefont {{Toon}}, \citenamefont {{Tran}}, \citenamefont {{Tyuterev}}, \citenamefont {{Adkins}}, \citenamefont {{Baker}}, \citenamefont {{Barbe}}, \citenamefont {{Can{\`e}}},
  \citenamefont {{Cs{\'a}sz{\'a}r}}, \citenamefont {{Dudaryonok}}, \citenamefont {{Egorov}}, \citenamefont {{Fleisher}}, \citenamefont {{Fleurbaey}}, \citenamefont {{Foltynowicz}}, \citenamefont {{Furtenbacher}}, \citenamefont {{Harrison}}, \citenamefont {{Hartmann}}, \citenamefont {{Horneman}}, \citenamefont {{Huang}}, \citenamefont {{Karman}}, \citenamefont {{Karns}}, \citenamefont {{Kassi}}, \citenamefont {{Kleiner}}, \citenamefont {{Kofman}}, \citenamefont {{Kwabia-Tchana}}, \citenamefont {{Lavrentieva}}, \citenamefont {{Lee}}, \citenamefont {{Long}}, \citenamefont {{Lukashevskaya}}, \citenamefont {{Lyulin}}, \citenamefont {{Makhnev}}, \citenamefont {{Matt}}, \citenamefont {{Massie}}, \citenamefont {{Melosso}}, \citenamefont {{Mikhailenko}}, \citenamefont {{Mondelain}}, \citenamefont {{M{\"u}ller}}, \citenamefont {{Naumenko}}, \citenamefont {{Perrin}}, \citenamefont {{Polyansky}}, \citenamefont {{Raddaoui}}, \citenamefont {{Raston}}, \citenamefont {{Reed}}, \citenamefont {{Rey}}, \citenamefont
  {{Richard}}, \citenamefont {{T{\'o}bi{\'a}s}}, \citenamefont {{Sadiek}}, \citenamefont {{Schwenke}}, \citenamefont {{Starikova}}, \citenamefont {{Sung}}, \citenamefont {{Tamassia}}, \citenamefont {{Tashkun}}, \citenamefont {{Vander Auwera}}, \citenamefont {{Vasilenko}}, \citenamefont {{Vigasin}}, \citenamefont {{Villanueva}}, \citenamefont {{Vispoel}}, \citenamefont {{Wagner}}, \citenamefont {{Yachmenev}},\ and\ \citenamefont {{Yurchenko}}}]{HITRAN}%
  \BibitemOpen
  \bibfield  {author} {\bibinfo {author} {\bibfnamefont {I.~E.}\ \bibnamefont {{Gordon}}}, \bibinfo {author} {\bibfnamefont {L.~S.}\ \bibnamefont {{Rothman}}}, \bibinfo {author} {\bibfnamefont {R.~J.}\ \bibnamefont {{Hargreaves}}}, \bibinfo {author} {\bibfnamefont {R.}~\bibnamefont {{Hashemi}}}, \bibinfo {author} {\bibfnamefont {E.~V.}\ \bibnamefont {{Karlovets}}}, \bibinfo {author} {\bibfnamefont {F.~M.}\ \bibnamefont {{Skinner}}}, \bibinfo {author} {\bibfnamefont {E.~K.}\ \bibnamefont {{Conway}}}, \bibinfo {author} {\bibfnamefont {C.}~\bibnamefont {{Hill}}}, \bibinfo {author} {\bibfnamefont {R.~V.}\ \bibnamefont {{Kochanov}}}, \bibinfo {author} {\bibfnamefont {Y.}~\bibnamefont {{Tan}}}, \bibinfo {author} {\bibfnamefont {P.}~\bibnamefont {{Wcis{\l}o}}}, \bibinfo {author} {\bibfnamefont {A.~A.}\ \bibnamefont {{Finenko}}}, \bibinfo {author} {\bibfnamefont {K.}~\bibnamefont {{Nelson}}}, \bibinfo {author} {\bibfnamefont {P.~F.}\ \bibnamefont {{Bernath}}}, \bibinfo {author} {\bibfnamefont {M.}~\bibnamefont
  {{Birk}}}, \bibinfo {author} {\bibfnamefont {V.}~\bibnamefont {{Boudon}}}, \bibinfo {author} {\bibfnamefont {A.}~\bibnamefont {{Campargue}}}, \bibinfo {author} {\bibfnamefont {K.~V.}\ \bibnamefont {{Chance}}}, \bibinfo {author} {\bibfnamefont {A.}~\bibnamefont {{Coustenis}}}, \bibinfo {author} {\bibfnamefont {B.~J.}\ \bibnamefont {{Drouin}}}, \bibinfo {author} {\bibfnamefont {J.~M.}\ \bibnamefont {{Flaud}}}, \bibinfo {author} {\bibfnamefont {R.~R.}\ \bibnamefont {{Gamache}}}, \bibinfo {author} {\bibfnamefont {J.~T.}\ \bibnamefont {{Hodges}}}, \bibinfo {author} {\bibfnamefont {D.}~\bibnamefont {{Jacquemart}}}, \bibinfo {author} {\bibfnamefont {E.~J.}\ \bibnamefont {{Mlawer}}}, \bibinfo {author} {\bibfnamefont {A.~V.}\ \bibnamefont {{Nikitin}}}, \bibinfo {author} {\bibfnamefont {V.~I.}\ \bibnamefont {{Perevalov}}}, \bibinfo {author} {\bibfnamefont {M.}~\bibnamefont {{Rotger}}}, \bibinfo {author} {\bibfnamefont {J.}~\bibnamefont {{Tennyson}}}, \bibinfo {author} {\bibfnamefont {G.~C.}\ \bibnamefont {{Toon}}},
  \bibinfo {author} {\bibfnamefont {H.}~\bibnamefont {{Tran}}}, \bibinfo {author} {\bibfnamefont {V.~G.}\ \bibnamefont {{Tyuterev}}}, \bibinfo {author} {\bibfnamefont {E.~M.}\ \bibnamefont {{Adkins}}}, \bibinfo {author} {\bibfnamefont {A.}~\bibnamefont {{Baker}}}, \bibinfo {author} {\bibfnamefont {A.}~\bibnamefont {{Barbe}}}, \bibinfo {author} {\bibfnamefont {E.}~\bibnamefont {{Can{\`e}}}}, \bibinfo {author} {\bibfnamefont {A.~G.}\ \bibnamefont {{Cs{\'a}sz{\'a}r}}}, \bibinfo {author} {\bibfnamefont {A.}~\bibnamefont {{Dudaryonok}}}, \bibinfo {author} {\bibfnamefont {O.}~\bibnamefont {{Egorov}}}, \bibinfo {author} {\bibfnamefont {A.~J.}\ \bibnamefont {{Fleisher}}}, \bibinfo {author} {\bibfnamefont {H.}~\bibnamefont {{Fleurbaey}}}, \bibinfo {author} {\bibfnamefont {A.}~\bibnamefont {{Foltynowicz}}}, \bibinfo {author} {\bibfnamefont {T.}~\bibnamefont {{Furtenbacher}}}, \bibinfo {author} {\bibfnamefont {J.~J.}\ \bibnamefont {{Harrison}}}, \bibinfo {author} {\bibfnamefont {J.~M.}\ \bibnamefont {{Hartmann}}},
  \bibinfo {author} {\bibfnamefont {V.~M.}\ \bibnamefont {{Horneman}}}, \bibinfo {author} {\bibfnamefont {X.}~\bibnamefont {{Huang}}}, \bibinfo {author} {\bibfnamefont {T.}~\bibnamefont {{Karman}}}, \bibinfo {author} {\bibfnamefont {J.}~\bibnamefont {{Karns}}}, \bibinfo {author} {\bibfnamefont {S.}~\bibnamefont {{Kassi}}}, \bibinfo {author} {\bibfnamefont {I.}~\bibnamefont {{Kleiner}}}, \bibinfo {author} {\bibfnamefont {V.}~\bibnamefont {{Kofman}}}, \bibinfo {author} {\bibfnamefont {F.}~\bibnamefont {{Kwabia-Tchana}}}, \bibinfo {author} {\bibfnamefont {N.~N.}\ \bibnamefont {{Lavrentieva}}}, \bibinfo {author} {\bibfnamefont {T.~J.}\ \bibnamefont {{Lee}}}, \bibinfo {author} {\bibfnamefont {D.~A.}\ \bibnamefont {{Long}}}, \bibinfo {author} {\bibfnamefont {A.~A.}\ \bibnamefont {{Lukashevskaya}}}, \bibinfo {author} {\bibfnamefont {O.~M.}\ \bibnamefont {{Lyulin}}}, \bibinfo {author} {\bibfnamefont {V.~Y.}\ \bibnamefont {{Makhnev}}}, \bibinfo {author} {\bibfnamefont {W.}~\bibnamefont {{Matt}}}, \bibinfo {author}
  {\bibfnamefont {S.~T.}\ \bibnamefont {{Massie}}}, \bibinfo {author} {\bibfnamefont {M.}~\bibnamefont {{Melosso}}}, \bibinfo {author} {\bibfnamefont {S.~N.}\ \bibnamefont {{Mikhailenko}}}, \bibinfo {author} {\bibfnamefont {D.}~\bibnamefont {{Mondelain}}}, \bibinfo {author} {\bibfnamefont {H.~S.~P.}\ \bibnamefont {{M{\"u}ller}}}, \bibinfo {author} {\bibfnamefont {O.~V.}\ \bibnamefont {{Naumenko}}}, \bibinfo {author} {\bibfnamefont {A.}~\bibnamefont {{Perrin}}}, \bibinfo {author} {\bibfnamefont {O.~L.}\ \bibnamefont {{Polyansky}}}, \bibinfo {author} {\bibfnamefont {E.}~\bibnamefont {{Raddaoui}}}, \bibinfo {author} {\bibfnamefont {P.~L.}\ \bibnamefont {{Raston}}}, \bibinfo {author} {\bibfnamefont {Z.~D.}\ \bibnamefont {{Reed}}}, \bibinfo {author} {\bibfnamefont {M.}~\bibnamefont {{Rey}}}, \bibinfo {author} {\bibfnamefont {C.}~\bibnamefont {{Richard}}}, \bibinfo {author} {\bibfnamefont {R.}~\bibnamefont {{T{\'o}bi{\'a}s}}}, \bibinfo {author} {\bibfnamefont {I.}~\bibnamefont {{Sadiek}}}, \bibinfo {author}
  {\bibfnamefont {D.~W.}\ \bibnamefont {{Schwenke}}}, \bibinfo {author} {\bibfnamefont {E.}~\bibnamefont {{Starikova}}}, \bibinfo {author} {\bibfnamefont {K.}~\bibnamefont {{Sung}}}, \bibinfo {author} {\bibfnamefont {F.}~\bibnamefont {{Tamassia}}}, \bibinfo {author} {\bibfnamefont {S.~A.}\ \bibnamefont {{Tashkun}}}, \bibinfo {author} {\bibfnamefont {J.}~\bibnamefont {{Vander Auwera}}}, \bibinfo {author} {\bibfnamefont {I.~A.}\ \bibnamefont {{Vasilenko}}}, \bibinfo {author} {\bibfnamefont {A.~A.}\ \bibnamefont {{Vigasin}}}, \bibinfo {author} {\bibfnamefont {G.~L.}\ \bibnamefont {{Villanueva}}}, \bibinfo {author} {\bibfnamefont {B.}~\bibnamefont {{Vispoel}}}, \bibinfo {author} {\bibfnamefont {G.}~\bibnamefont {{Wagner}}}, \bibinfo {author} {\bibfnamefont {A.}~\bibnamefont {{Yachmenev}}},\ and\ \bibinfo {author} {\bibfnamefont {S.~N.}\ \bibnamefont {{Yurchenko}}},\ }\href {https://doi.org/10.1016/j.jqsrt.2021.107949} {\ \textbf {\bibinfo {volume} {277}},\ \bibinfo {eid} {107949} (\bibinfo {year}
  {2022})}\BibitemShut {NoStop}%
\bibitem [{\citenamefont {Gray}\ \emph {et~al.}(2025)\citenamefont {Gray}, \citenamefont {Sekine}, \citenamefont {Shen}, \citenamefont {Zacharias}, \citenamefont {Williams}, \citenamefont {Zhou}, \citenamefont {Chawlani}, \citenamefont {Ledezma}, \citenamefont {Englebert},\ and\ \citenamefont {Marandi}}]{gray2025}%
  \BibitemOpen
  \bibfield  {author} {\bibinfo {author} {\bibfnamefont {R.~M.}\ \bibnamefont {Gray}}, \bibinfo {author} {\bibfnamefont {R.}~\bibnamefont {Sekine}}, \bibinfo {author} {\bibfnamefont {M.}~\bibnamefont {Shen}}, \bibinfo {author} {\bibfnamefont {T.}~\bibnamefont {Zacharias}}, \bibinfo {author} {\bibfnamefont {J.}~\bibnamefont {Williams}}, \bibinfo {author} {\bibfnamefont {S.}~\bibnamefont {Zhou}}, \bibinfo {author} {\bibfnamefont {R.}~\bibnamefont {Chawlani}}, \bibinfo {author} {\bibfnamefont {L.}~\bibnamefont {Ledezma}}, \bibinfo {author} {\bibfnamefont {N.}~\bibnamefont {Englebert}},\ and\ \bibinfo {author} {\bibfnamefont {A.}~\bibnamefont {Marandi}},\ }\href {https://arxiv.org/abs/2501.15381} {\bibinfo {title} {Two-optical-cycle pulses from nanophotonic two-color soliton compression}} (\bibinfo {year} {2025}),\ \Eprint {https://arxiv.org/abs/2501.15381} {arXiv:2501.15381 [physics.optics]} \BibitemShut {NoStop}%
\end{thebibliography}
